\documentclass[a4paper,12pt]{article}

\usepackage{graphicx}
\usepackage{amsmath}
\usepackage{amssymb}
\usepackage{hyperref}
\usepackage{tabularx}
\newcolumntype{M}{>{$}c<{$}}
\usepackage[matrix,arrow,curve]{xy}

\numberwithin{equation}{section} \numberwithin{figure}{section}
\numberwithin{table}{section}

\def\papertitlepage{\baselineskip 3.5ex\thispagestyle{empty}}
\def\Title#1{\baselineskip 1cm \vspace{1.5cm}%
  \begin{center}{\Large\bf #1}\end{center}\vspace{0.5cm}}
\def\Authors#1{\begin{center}\renewcommand{\thefootnote}{\fnsymbol{footnote}}{\it #1}\end{center}}
\def\Abstract{\vspace{1.0cm}%
  \begin{center}{\large\bf Abstract}\end{center}}

\renewenvironment{thebibliography}{\pagebreak[3]\par\vspace{0.6em}
\begin{flushleft}{\large \bf References}\end{flushleft}
\vspace{-1.0em}

\begin{enumerate}\if@twocolumn\baselineskip=0.6em\itemsep -0.2em
\else\itemsep -0.2em\fi\labelsep 0.1em}{\end{enumerate} }

\DeclareMathDelimiter{\lcolon}{\mathopen}{operators}{"3A}{largesymbols}{"3A}
\DeclareMathDelimiter{\rcolon}{\mathclose}{operators}{"3A}{largesymbols}{"3A}

\def\+{\!\!+\!\!}

\def\dynkin(#1){(#1)}

\def\bra<#1|{\langle#1|}
\def\ket|#1>{|#1\rangle}
\def\braket<#1|#2>{\langle#1|#2\rangle}
\def\llangle{\langle\!\langle}
\def\rrangle{\rangle\!\rangle}
\def\bbra<#1|{\llangle#1|}
\def\kket|#1>{|#1\rrangle}
\def\bbraket<#1|#2>{\llangle#1|#2\rrangle}

\begin{document}
{\papertitlepage \vspace*{0cm} {\hfill
\begin{minipage}{4.2cm}
IF-USP 2011\par\noindent June, 2011
\end{minipage}}
\Title{Conservation laws and tachyon potentials in the sliver
frame}
\Authors{{\sc E.~Aldo~Arroyo${}$\footnote{\tt
aldohep@fma.if.usp.br}}
\\
Instituto de F\'{i}sica, Universidade de S\~{a}o Paulo \\[-2ex]
C.P. 66.318 CEP 05315-970, S\~{a}o Paulo, SP, Brasil ${}$ }
} 

\vskip-\baselineskip
{\baselineskip .5cm \Abstract Conservation laws have provided an
elegant and efficient tool to evaluate the open string field
theory interaction vertex, they have been originally implemented
in the case where the string field is expanded in the Virasoro
basis. In this work we derive conservation laws in the case where
the string field is expanded in the so-called sliver
$\mathcal{L}_0$-basis. As an application of these conservation
laws derived in the sliver frame, we compute the open string field
action relevant to the tachyon condensation and in order to
present not only an illustration but also an additional
information, we evaluate the action without imposing a gauge
choice.

\newpage
\setcounter{footnote}{0}
\tableofcontents

\section{Introduction}
Since Schnabl's discovery of the first analytic solution for
tachyon condensation \cite{Schnabl:2005gv}, in the last few years
string field theory has come to the surface as a suitable
framework to verify analytically Sen's conjectures
\cite{Sen:1999mh,Sen:1999xm}. Schnabl's solution successfully
proves Sen's first conjecture which states that at the stationary
point of the tachyon potential on a D25-brane of open bosonic
string theory, the negative energy density exactly cancels the
tension of the D25-brane. The second conjecture states that there
are lump solutions which describe lower dimensional D-branes; and
the third conjecture tell us that in the nonperturbative tachyon
vacuum all open string degrees of freedom must disappear
\footnote{There are similar conjectures for the open string
tachyon on a non-BPS D-brane and the tachyon living on the
brane-antibrane pair
\cite{Sen:1998ii,Sen:1998sm,Sen:1999mg,Bai:2005jr,Bagchi:2008et,Garousi:2007fk,Garousi:2008ge}.}.
Witten's formulation of open bosonic string field theory
\cite{Witten:1985cc} has been proven to be the suitable framework
to test Sen's conjectures
\cite{Kostelecky:1988ta,Kostelecky:1989nt,Kostelecky:1995qk,Sen:1999nx,Moeller:2000xv,Taylor:2002fy,Gaiotto:2002wy,
Harvey:2000tv,de Mello
Koch:2000ie,Moeller:2000jy,Ellwood:2001py,Ellwood:2001ig,Giusto:2003wc,Ellwood:2006ba,Kwon:2007mh,AldoArroyo:2009hf,Arroyo:2010sy,Arroyo:2010fq,
Okawa:2006vm,Fuchs:2006hw,Takahashi:2007du,Aref'eva:2009ac,Arroyo:2009ec,Erler:2009uj,Zeze:2010jv,Ellwood:2008jh,Kawano:2008ry,Kawano:2008jv,Kiermaier:2008qu,
Rastelli:2001jb,Schnabl:2010tb,Murata:2011ex,Bonora:2010hi,Kiermaier:2010cf}.

Historically the tachyonic instability in Witten's cubic open
string field theory has been analyzed by looking at the extremum
of the tachyon potential at which the total negative potential
energy exactly cancels the tension of the D25-brane. The tachyon
potential has been computed numerically using the level truncation
scheme
\cite{Kostelecky:1988ta,Kostelecky:1989nt,Kostelecky:1995qk,Sen:1999nx,Moeller:2000xv,Taylor:2002fy,Gaiotto:2002wy}.
This scheme originally introduced by Kostelecky and Samuel is
based on the realization that by truncating the string field to
its low lying modes, namely keeping only Fock states with $L_0$
eigenvalue less than $h$, one obtains an approximation that gets
more accurate as the level $h$ is increased. Therefore in order to
perform level truncation computations, the string field was
traditionally expanded in the usual Virasoro basis of $L_0$
eigenstates. However, it is well known that in this basis
calculations involving the cubic interaction term becomes
cumbersome and the three-string vertex that defines the star
product in the string field algebra is complicated
\cite{Gross:1986ia,Gross:1986fk,LeClair:1988sp,LeClair:1988sj}. We
can overcome these technical issues (related to the definition of
the star product) by using a new coordinate system
\cite{Schnabl:2005gv}.

Let us recall that the open string worldsheet is usually
parameterized by a complex strip coordinate $w=\sigma + i \tau$,
$\sigma \in [0,\pi]$, or by $z=-e^{-iw}$, which takes values on
the upper half plane. As shown in \cite{Schnabl:2005gv}, the
gluing conditions entering into the geometrical definition of the
star product simplify if one uses another coordinate system,
$\tilde z = \arctan z$, in which the upper half plane looks as a
semi-infinite cylinder of circumference $\pi$. In this new
coordinate system, which we refer as the \textit{sliver frame}, it
is possible to write down simple, closed expressions for arbitrary
star products within the subalgebra generated by Fock space
states. Elements of this subalgebra are known in the literature
\cite{Rastelli:2000iu,David:2001xu,Schnabl:2002gg} as wedge states
with insertions.

Using this new coordinate system, Schnabl's analytic solution of
the open bosonic string field theory equations of motion was found
by expanding the string field in a basis of ${\cal L}_0$
eigenstates \cite{Schnabl:2005gv}. The operator ${\cal L}_0$ is
the zero mode of the worldsheet energy momentum tensor in the
sliver frame. By a conformal transformation it can be written as
\begin{eqnarray}
\mathcal{L}_0= \oint \frac{d z}{2 \pi i} (1+z^2) \arctan z \,
T(z)=L_0+\sum_{k=1}^{\infty} \frac{2(-1)^{k+1}}{4k^2-1}L_{2k} \; ,
\end{eqnarray}
where the $L_n$'s are the ordinary Virasoro generators with zero
central charge $c=0$ of the total matter and ghost conformal field
theory.

Since in open bosonic string field theory, the basic field is an
object of ghost number one, a rather generic string field $\Psi$
can be written as
\begin{eqnarray}
\label{ansa01ss} \Psi=\sum_{n,p} f_{n,p} \hat{\mathcal{L}}^n
\tilde{c}_p |0\rangle +\sum_{n,p,q} f_{n,p,q}
\hat{\mathcal{B}}\hat{\mathcal{L}}^n \tilde{c}_p \tilde{c}_q
|0\rangle \, ,
\end{eqnarray}
where $n = 0, 1, 2, \cdots$, and $p, q = 1, 0,-1,-2, \cdots \cdot$
The operators $\hat{\mathcal{L}}$, $\hat{\mathcal{B}}$ and $\tilde
c_p$ are defined in the sliver frame \cite{Schnabl:2005gv}, and
they are related to the worldsheet energy momentum tensor, the $b$
and $c$ ghosts fields respectively
\begin{align}
\label{lhat1intro} \hat{\mathcal{L}}\equiv {\cal L}_0+{\cal
L}_0^\dag&= \oint \frac{d z}{2 \pi i} (1+z^{2}) (\arctan
z+\text{arccot} z) \, T(z) \, , \\
\label{bhat2intro} \hat{\mathcal{B}}\equiv {\cal B}_0+{\cal
B}_0^\dag&= \oint \frac{d z}{2 \pi i} (1+z^{2}) (\arctan
z+\text{arccot} z) \, b(z) \, , \\
 \tilde c_p & = \oint \frac{d z}{2 \pi i} \frac{1}{(1+z^{2})^{2}} (\arctan
z)^{p-2} \, c(z) \, .
\end{align}

To find an analytic solution which represents to the tachyon
vacuum, we impose some gauge and plug the string field
(\ref{ansa01ss}) into the equation of motion $Q_B \Psi + \Psi*\Psi
= 0$ so that the coefficients $f_{n,p}$ and $f_{n,p,q}$ can be
determined level by level. For example, imposing the gauge ${\cal
B}_0 \Psi=0$, Schnabl was able to determine these coefficients in
terms of the celebrated Bernoulli numbers \cite{Schnabl:2005gv}.
Once we have an analytic solution, in order to compute the value
of the vacuum energy, we replace the solution $\Psi$ into the
string field theory action
\begin{eqnarray}
\label{accionx1} S=-\Big[ \frac{1}{2} \langle \Psi,Q_B \Psi
\rangle +\frac{1}{3} \langle \Psi,\Psi*\Psi \rangle \Big] \, .
\end{eqnarray}

At this stage it is clear that to evaluate the vacuum energy, we
will need to compute correlation functions involving the two and
three point interaction vertex. For vertex operators given in
terms of the basic $\hat{\mathcal{L}}$, $\hat{\mathcal{B}}$, and
$\tilde c_p$ operators, to compute these correlation functions,
straightforward techniques have been developed. As it is shown in
\cite{AldoArroyo:2009hf}, correlation functions involving the two
point vertex, namely the BPZ inner product, can be expressed in
terms of a known function $\mathcal{F}$ evaluated at some
particular points. On the other hand correlators involving the
three point interaction vertex, namely correlators involving the
$*$-product can be expressed in terms of triple contour integrals
having as an integrand the function $\mathcal{F}$. In reference
\cite{AldoArroyo:2009hf}, we dare to compute explicitly these
triple contour integrals, but without any succeed, so we have
written a computer program to do it for us. However there is a
technical computational issue related to the evaluation of these
triple contour integrals. Since the number of states involved in
the $\mathcal{L}_0$ level expansion of the string field $\Psi$
grows rapidly with the level, the complexity of performing many
triple contour integrals, involved in the evaluation of the three
point interaction vertex $\langle \Psi, \Psi*\Psi \rangle$,
increases exponentially with the level. To solve this technical
issue, it would be nice to find an alternative and efficient
method to evaluate correlators involving the $*$-product.

Apart from the technicality commented above, we also would like to
mention that, in reference \cite{AldoArroyo:2009hf}, we have
derived expressions for correlation functions involving operators
of the form $\hat{\mathcal{L}}^n \tilde c_p |0 \rangle$ and
$\hat{\mathcal{L}}^n \hat{\mathcal{B}} \tilde c_p \tilde c_q|0
\rangle$. However we can have more general set of states, like the
ones discussed in Schnabl's original paper \cite{Schnabl:2005gv},
for instance states of the form
\begin{align}
\label{ba01i} \hat{\mathcal{L}}^m \mathcal{L}_{n_1} \cdots
\mathcal{L}_{n_k} \tilde c_p |0 \rangle \; &, \;\;\; \text{with}
\;\;\; n_1,\cdots,n_k =-2,-3,\cdots \\
\label{ba02i} \hat{\mathcal{L}}^m \hat{\mathcal{B}}
\mathcal{L}_{n_1} \cdots \mathcal{L}_{n_k} \tilde c_p \tilde c_q
|0 \rangle \; &, \;\;\; \text{with} \;\;\; n_1,\cdots,n_k
=-2,-3,\cdots \\
\label{ba03i} \hat{\mathcal{L}}^m \mathcal{L}_{n_1} \cdots
\mathcal{L}_{n_k} \mathcal{B}_{l} \tilde c_p \tilde c_q |0 \rangle
\; &, \;\;\; \text{with}
\;\;\; n_1,\cdots,n_k,l =-2,-3,\cdots \\
\label{ba04i} \hat{\mathcal{L}}^m \hat{\mathcal{B}}
\mathcal{L}_{n_1} \cdots \mathcal{L}_{n_k}  \mathcal{B}_{l} \tilde
c_p \tilde c_q \tilde c_r |0 \rangle \; & , \;\;\; \text{with}
\;\;\; n_1,\cdots,n_k,l =-2,-3,\cdots \;
\end{align}
where the operators $\mathcal{L}_{n_k}$ and $\mathcal{B}_{l}$ are
modes of the worldsheet energy momentum tensor and the $b$ ghost
in the $\tilde z$ coordinate. By a conformal transformation they
can be written as
\begin{eqnarray}
\label{llmi} \mathcal{L}_{n_k} &=& \oint \frac{d z}{2 \pi i}
(1+z^2)
(\arctan z)^{n_k+1} \, T(z) \, ,\\
\label{llmi} \mathcal{B}_{l} &=& \oint \frac{d z}{2 \pi i} (1+z^2)
(\arctan z)^{l+1} \, b(z) \, .
\end{eqnarray}

At this point we would like to address the question: is there an
alternative method for evaluating the three point interaction
vertex which involves generic vertex operators like the ones
discussed previously? We will show that the answer to this
question turns out to be affirmative. This kind of problem also
happens in the case where the string field is expanded in the
usual Virasoro basis. In that case there are two ways of computing
the interaction vertex: the first one is by means of finite
conformal transformations of the three vertex operators, the
second and more efficient method is by using conservation laws
\cite{Rastelli:2000iu}.

In order to implement this alternative method to compute the three
point interaction vertex, in this paper we derive conservation
laws in the case where the string field is expanded in the sliver
$\mathcal{L}_0$-basis. Schematically, in this method based on
conservation laws, the task consists in finding (inside the three
point correlation function) an expression for the operators
$\hat{\mathcal{L}}^m$, $\hat{\mathcal{B}}$, $\mathcal{L}_{n_k}$
and $\mathcal{B}_{l}$ (with $n_k,l =-2,-3,\cdots$) in terms of the
modes $\mathcal{L}_{u}$ and $\mathcal{B}_{v}$ with $u,v
=-1,0,1,\cdots$. Since for these modes we have $\mathcal{L}_{u} |0
\rangle=0$, $\mathcal{B}_{v} |0 \rangle=0$, the commutation and
anticommutation relations of $\mathcal{L}_{u}$ and $\mathcal{B}_v$
with the rest of operators allow to find a recursive procedure
that express the three point correlation function of any state in
terms of correlators which only involve modes of the $c$ ghost. As
an application of this new set of conservation laws, we compute
the open string field action relevant to the tachyon condensation
and in order to present not only an illustration but also an
additional information, we evaluate the action without imposing a
gauge choice.

This paper is organized as follows. In section 2, we review the
straightforward techniques employed in the evaluation of
correlation functions for operators defined in the sliver frame.
In section 3, we discuss in detail the derivation of conservation
laws for operators which are widely used in the $\mathcal{L}_0$
level expansion of the string field. In section 4, as an
application of the conservation laws derived in the previous
section, we compute the open string field action relevant to the
tachyon condensation and in order to present not only an
illustration but also an additional information, we evaluate the
action without imposing a gauge choice. In section 5, a summary
and further directions of exploration are given.

\section{Correlation functions in the sliver frame}
In this section, we will review the straightforward techniques
used in the computation of correlation functions of operators
defined in the sliver frame. Let us remark that the results shown
in this section were derived in reference
\cite{AldoArroyo:2009hf}. We begin with the description of the two
and three string vertex which are used in the definition of the
kinetic and cubic term of the open string field action.

\subsection{The two and three string vertex}
The relation between a point in the upper half plane (UHP) $z$ and
a point in the sliver frame $\tilde z$ is given by the conformal
map $\tilde z = \arctan z$. Note that by this conformal
transformation, the UHP is mapped to a semi-infinite cylinder
$C_{\pi}$ of circumference $\pi$. It turns out that the sliver
frame seems to be the most natural one since the conformal field
theory in this new coordinate system remains easy. As in the case
of the UHP, we can define general $n$-point correlation functions
on $C_\pi$ which can be readily found in terms of correlation
functions defined on the UHP by a conformal mapping,
\begin{eqnarray}
\langle \phi_1(\tilde x_1) \cdots \phi_n(\tilde
x_n)\rangle_{C_\pi}=\langle\tilde\phi_1(\tilde x_1) \cdots
\tilde\phi_n(\tilde x_n)\rangle_{UHP} \; ,
\end{eqnarray}
where the fields $\tilde\phi_i(\tilde x_i)$ are defined as
conformal transformation $\tilde\phi_i(\tilde x_i)= \tan \circ \,
\phi_i(\tilde x_i)$. In general $f \circ \mathcal{\phi}$ denotes a
conformal transformation of a field $\phi$ under a map $f$, for
instance if $\phi$ represents a primary field of dimension $h$,
then $f \circ \mathcal{\phi}$ is defined as $f \circ
\mathcal{\phi}(x)=(f'(x))^h \phi(f(x))$.

The two-string vertex which appears in the string field theory action is
the familiar BPZ inner product of conformal field theory
\footnote{Recall that the BPZ conjugate for the modes of an
holomorphic field $\phi$ of dimension $h$ is given by
bpz$(\phi_n)=(-1)^{n+h}\phi_{-n}$.}. It is defined as a map
$\mathcal{H} \otimes \mathcal{H} \rightarrow \mathbb{R}$
\begin{eqnarray}
\langle\phi_1,\phi_2\rangle = \langle \mathcal{I} \circ \,
\phi_1(0) \phi_2(0) \rangle_{UHP} \; ,
\end{eqnarray}
where $\mathcal{I}: z \rightarrow -1/z$ is the inversion symmetry.
For states defined on the sliver frame $|\tilde\phi_i>$ the
two-string vertex can be written as
\begin{eqnarray}
\label{twooo2} \langle \tilde\phi_1,\tilde\phi_2\rangle = \langle
\mathcal{I} \circ \, \tilde \phi_1(0) \tilde
\phi_2(0)\rangle_{UHP}=\langle\phi_1(\frac{\pi}{2})
\phi_2(0)\rangle_{C_{\pi}} \, .
\end{eqnarray}
As we can see in this last expression, we evaluate the two-string vertex
at two different points, namely at $\pi/2$ and $0$ on $C_{\pi}$. This must be the case since the inversion symmetry maps the point
at $z=0$ on the upper half plane to the point at infinity,
but the point at infinity is mapped to the point $\pm \pi/2$ on
$C_{\pi}$.

The three-string vertex is a map $\mathcal{H} \otimes
\mathcal{H} \otimes \mathcal{H} \rightarrow \mathbb{R}$, and it is
defined as a correlator on a surface formed by gluing together three
strips representing three open strings. For states defined on the
sliver frame $|\tilde\phi_i>$ the three-string vertex can be written as
\begin{eqnarray}
\label{tr333} \langle\tilde\phi_1,\tilde\phi_2,\tilde\phi_3\rangle
= \langle \phi_1(\frac{3 \pi}{4}) \phi_2(\frac{\pi}{4})
\phi_3(-\frac{\pi}{4})\rangle_{C_{   \frac{3 \pi}{2}   }} \, .
\end{eqnarray}
Here the correlator is taken on a semi-infinite cylinder $C_{
\frac{3 \pi}{2}}$ of circumference $ 3 \pi/2$. Also, this
correlator can be evaluated on the semi-infinite cylinder $C_\pi$
of circumference $\pi$. We only need to perform a simple conformal
map (scaling) $s:\tilde z \rightarrow \frac{2}{3} \tilde z$ which
brings the region $C_{ \frac{3 \pi}{2}}$ to $C_\pi$, and the
correlator is given by
\begin{eqnarray}
\langle\tilde\phi_1,\tilde\phi_2,\tilde\phi_3\rangle = \langle s
\circ \, \phi_1(\frac{3 \pi}{4}) s \circ \, \phi_2(\frac{\pi}{4})
s \circ \, \phi_3(-\frac{\pi}{4})\rangle_{C_\pi} \, .
\end{eqnarray}
Note that the scaling transformation $s$ is implemented by $U_3 =
(2/3)^{\mathcal{L}_0}$, where $\mathcal{L}_0$ is the zero mode of
the worldsheet energy momentum tensor $T_{\tilde z \tilde
z}(\tilde z)$ in the $\tilde z$ coordinate,
\begin{eqnarray}
\mathcal{L}_0=\oint \frac{d \tilde z}{2 \pi i} \tilde z T_{\tilde
z \tilde z}(\tilde z) \; .
\end{eqnarray}
By a conformal transformation it can be expressed as
\begin{eqnarray}
\mathcal{L}_0= \oint \frac{d z}{2 \pi i} (1+z^2) \arctan z T_{ z
z}(z)=L_0+\sum_{k=1}^{\infty} \frac{2(-1)^{k+1}}{4k^2-1}L_{2k} \;
,
\end{eqnarray}
where the $L_n$'s are the ordinary Virasoro generators (with zero
central charge) of the full (matter plus ghost) conformal field
theory.

\subsection{Evaluating some correlators}
Using the definition of the conformal transformation $\tilde
c(x)=\cos^2(x) c(\tan x)$ of the $c$ ghost and its anticommutation
relations with the operators $Q_B$, $\mathcal{B}_0$ and $B_1$,
\footnote{The operators $\mathcal{B}_0$ and
$B_1\equiv\mathcal{B}_{-1}$ are modes of the $b$ ghost which are
defined on the semi-infinite cylinder coordinate as
$\mathcal{B}_{n}=\oint \frac{dz}{2 \pi i}(1+z^2) (\arctan
z)^{n+1}b(z)$.}
\begin{align}
\{Q_B,\tilde c(z)\}&= \tilde c(z) \partial \tilde c(z) \, , \\
\{\mathcal{B}_0,\tilde c(z)\}&=z \, , \\
\{B_1,\tilde c(z)\}&=1 \, ,
\end{align}
we obtain the following basic correlation functions,
\begin{align}
\label{a0} \langle \tilde{c}(x)\tilde{c}(y)\tilde{c}(z) \rangle
&=\sin(x -y)\sin(x - z)\sin(y - z) \, , \\
\label{a1} \langle \tilde c(x) Q_B \tilde c(y) \rangle &= - \sin(x-y)^2 \, , \\
\label{a2}\langle \tilde{c}(x)\mathcal{B}_0
\tilde{c}(y)\tilde{c}(z) \tilde{c}(w)\rangle &= y \langle
\tilde{c}(x)\tilde{c}(z)\tilde{c}(w)\rangle-z \langle
\tilde{c}(x)\tilde{c}(y)\tilde{c}(w)\rangle +w \langle
\tilde{c}(x)\tilde{c}(y)\tilde{c}(z) \rangle \, , \\
\label{a3}\langle \tilde{c}(x)\tilde{c}(y)\mathcal{B}_0
\tilde{c}(z) \tilde{c}(w)\rangle &= z \langle
\tilde{c}(x)\tilde{c}(y)\tilde{c}(w)\rangle -w \langle
\tilde{c}(x)\tilde{c}(y)\tilde{c}(z) \rangle \, , \\
\label{a4}\langle \tilde{c}(x)B_1 \tilde{c}(y)\tilde{c}(z)
\tilde{c}(w)\rangle &= \langle
\tilde{c}(x)\tilde{c}(z)\tilde{c}(w)\rangle-\langle
\tilde{c}(x)\tilde{c}(y)\tilde{c}(w)\rangle + \langle
\tilde{c}(x)\tilde{c}(y)\tilde{c}(z) \rangle \, , \\
\label{a5}\langle \tilde{c}(x)\tilde{c}(y)B_1 \tilde{c}(z)
\tilde{c}(w)\rangle &= \langle
\tilde{c}(x)\tilde{c}(y)\tilde{c}(w)\rangle - \langle
\tilde{c}(x)\tilde{c}(y)\tilde{c}(z) \rangle \, .
\end{align}

In order to write, in a short way, long and messy expressions, the
following definitions will be very useful
\begin{align}
\label{sa}\sigma(a)&\equiv \oint \frac{dz}{2 \pi i} z^{a} \sin(2z) \nonumber \\
&= \frac{\theta(-a-2)}{\Gamma(-a)} ((-1)^{a}+1)
(-1)^{\frac{2-a}{2}} 2^{-a-2} \, , \\
\label{ca}\varsigma(a)&\equiv \oint \frac{dz}{2 \pi i} z^{a} \cos(2z) \nonumber \\
&= \frac{\theta(-a-1)}{\Gamma(-a)} ((-1)^{a}-1)
(-1)^{\frac{1-a}{2}} 2^{-a-2} \, ,
\end{align}
\begin{align}
\label{f}
\mathcal{F}(a_1,a_2,a_3,\alpha_1,\beta_1,\alpha_2,\beta_2,\alpha_3,\beta_3)\equiv
\oint \frac{dx_1 dx_2 dx_3}{(2 \pi i)^3}
x_1^{a_1}x_2^{a_2}x_3^{a_3} \langle \tilde c(\alpha_1 x_1
+\beta_1) \tilde c(\alpha_2 x_2 +\beta_2)\tilde c(\alpha_3 x_3
+\beta_3) \rangle \nonumber \;\;\;\;\;\;\;\;\;\;\\
= \frac{1}{ \alpha_1^{a_1+1}\alpha_2^{a_2+1}\alpha_3^{a_3+1}}
\Big[
\;\;\;\;\;\;\;\;\;\;\;\;\;\;\;\;\;\;\;\;\;\;\;\;\;\;\;\;\;\;\;\;\;\;\;\;\;\;\;\;\;\;\;\;\;\;\;\;\;\;\;\;\;\;
 \;\;\;\;\;\;\;\;\;\;\;\;\;\;\;\;\;\;\;\;\;\;\;\;   \nonumber\\
 \delta_{a_3,-1}\frac{ \big(\sigma(a_1)\sigma(a_2)+\varsigma(a_1)\varsigma(a_2)\big) \sin(2(\beta_1-\beta_2))
 +
 \big(\sigma(a_1)\varsigma(a_2)-\varsigma(a_1)\sigma(a_2)\big) \cos(2(\beta_1-\beta_2))}{4} \nonumber \;\;\;\;\;\;\;\;\;\;\;\;\;\;\;\;\;\;\;\;\\
 + \delta_{a_2,-1}\frac{ \big(\varsigma(a_1)\sigma(a_3)-\sigma(a_1)\varsigma(a_3)\big) \cos(2(\beta_1-\beta_3))
 -
 \big(\varsigma(a_1)\varsigma(a_3)+\sigma(a_1)\sigma(a_3)\big) \sin(2(\beta_1-\beta_3))}{4} \nonumber \;\;\;\;\;\;\;\;\;\;\;\;\;\;\;\;\;\;\\
 + \delta_{a_1,-1}\frac{ \big(\varsigma(a_2)\mathcal{\varsigma}(a_3)+\sigma(a_2)\sigma(a_3)\big) \sin(2(\beta_2-\beta_3))
 +
 \big(\sigma(a_2)\varsigma(a_3)-\varsigma(a_2)\sigma(a_3)\big)
 \cos(2(\beta_2-\beta_3))}{4}
 \Big], \;\;\;\;\;\;\;\;\;\;\;\;\;\;\; \nonumber \\
\end{align}
where $\theta(n)$ is the unit step (Heaviside) function which is
defined as
\begin{align}
\theta(n) =
\begin{cases}
  0,  & \mbox{if }n < 0 \\
  1, & \mbox{if }n \geq 0 \, .
\end{cases}
\end{align}

Let us list some correlation functions involving operators
frequently used in the $\mathcal{L}_0$-basis, namely the operators
$\hat{\mathcal{L}}^{n}$ ($\hat{\mathcal{L}}\equiv {\cal L}_0+{\cal
L}_0^\dag$), $\hat{\mathcal{B}}$ ($\hat{\mathcal{B}}\equiv {\cal
B}_0+{\cal B}_0^\dag$), $U_r=\big(\frac{2}{r}\big)^{{\cal L}_0}$
and the $\tilde c(z)$ ghost
\begin{align}
\label{correla1}&\langle \text{bpz}(\tilde{c}_{p_1})
\hat{\mathcal{L}}^{n_1} U^\dag_{r} U_{r}
\tilde{c}(x)\tilde{c}(y) \rangle = \nonumber \\
&= \oint \frac{dz_1 dx_1}{(2 \pi i)^{2}}\frac{(-2)^{n_1} n_1! \,
x_1^{p_1-2}}{(z_1-2)^{n_1+1}} \big(\frac{2}{r}\big)^{-p_1+n_1-2}
\big(\frac{2}{z_1}\big)^{-p_1-2} \langle
\tilde{c}(x_1+\frac{\pi}{2}) \tilde{c}(\frac{4}{z_1
r}x)\tilde{c}(\frac{4}{z_1 r}y) \rangle \, ,
\end{align}
\begin{align}
\label{correla2}&\langle \text{bpz}(\tilde{c}_{p_1})
\hat{\mathcal{L}}^{n_1} \hat{\mathcal{B}} U^\dag_{r} U_{r}
\tilde{c}(x)\tilde{c}(y)\tilde{c}(z) \rangle = \nonumber \\
&= - \delta_{p_1,0} \oint \frac{dz_1}{ 2 \pi i} \frac{(-2)^{n_1}
n_1!}{(z_1-2)^{n_1+1}} \big(\frac{2}{r}\big)^{-p_1+n_1-2}
\big(\frac{2}{z_1}\big)^{-p_1-2} \langle \tilde{c}(\frac{4}{z_1
r}x)\tilde{c}(\frac{4}{z_1 r}y)\tilde{c}(\frac{4}{z_1 r}z) \rangle
\nonumber \\
&+  \oint \frac{dz_1 dx_1}{(2 \pi i)^{2}}\frac{(-2)^{n_1} n_1! \,
x_1^{p_1-2}}{(z_1-2)^{n_1+1}} \big(\frac{2}{r}\big)^{-p_1+n_1-2}
\big(\frac{2}{z_1}\big)^{-p_1-2} \langle
\tilde{c}(x_1+\frac{\pi}{2}) \mathcal{B}_0 \tilde{c}(\frac{4}{z_1
r}x)\tilde{c}(\frac{4}{z_1 r}y)\tilde{c}(\frac{4}{z_1 r}z) \rangle
\, ,
\end{align}
\begin{align}
\label{correla3}&\langle \text{bpz}(\tilde{c}_{p_1})
\text{bpz}(\tilde{c}_{p_2}) \hat{\mathcal{L}}^{n_1}
\hat{\mathcal{B}} U^\dag_{r} U_{r}
\tilde{c}(x)\tilde{c}(y) \rangle = \nonumber \\
&= - \delta_{p_2,0}\oint \frac{dz_1 dx_1}{(2 \pi
i)^{2}}\frac{(-2)^{n_1} n_1! \, x_1^{p_1-2}}{(z_1-2)^{n_1+1}}
\big(\frac{2}{r}\big)^{-p_1-p_2+n_1-1}
\big(\frac{2}{z_1}\big)^{-p_1-p_2-1} \langle
\tilde{c}(x_1+\frac{\pi}{2}) \tilde{c}(\frac{4}{z_1
r}x)\tilde{c}(\frac{4}{z_1 r}y) \rangle \nonumber \\
& +\delta_{p_1,0}\oint \frac{dz_1 dx_2}{(2 \pi
i)^{2}}\frac{(-2)^{n_1} n_1! \, x_2^{p_2-2}}{(z_1-2)^{n_1+1}}
\big(\frac{2}{r}\big)^{-p_1-p_2+n_1-1}
\big(\frac{2}{z_1}\big)^{-p_1-p_2-1} \langle
\tilde{c}(x_2+\frac{\pi}{2}) \tilde{c}(\frac{4}{z_1
r}x)\tilde{c}(\frac{4}{z_1 r}y) \rangle \nonumber \\
& +\oint \frac{dz_1 dx_1 dx_2}{(2 \pi i)^{3}}\frac{(-2)^{n_1} n_1!
\, x_1^{p_1-2} x_2^{p_2-2}}{(z_1-2)^{n_1+1}}
\big(\frac{2}{r}\big)^{-p_1-p_2+n_1-1}
\big(\frac{2}{z_1}\big)^{-p_1-p_2-1} \times \nonumber \\
 &\;\;\;\;\;\;\;\;\;\;\;\;\;\;\;\;\;\;\;\;\;\;\;\;\;\;\;\;
 \;\;\;\;\;\;\;\;\;\;\;\;\;\;\;\;\;\;\;\;\;\;\;
 \;\;\;\;\;\;\;\;\;\;\;\;\;\;\;\;\;\; \times \langle \tilde{c}(x_1+\frac{\pi}{2})
\tilde{c}(x_2+\frac{\pi}{2}) \mathcal{B}_0 \tilde{c}(\frac{4}{z_1
r}x)\tilde{c}(\frac{4}{z_1 r}y) \rangle \, ,
\end{align}
where the \emph{bpz} acting on the modes of the $\tilde c(z)$
ghost stands for the usual BPZ conjugation which in the ${\cal
L}_0$-basis is defined as follows
\begin{align}
\text{bpz}(\tilde \phi_n) &= \oint \frac{d  z}{2 \pi i} z^{n+h-1}
\tilde \phi ( z + \frac{\pi}{2}) \, ,
\end{align}
for any primary field $\tilde \phi(z)$ with weight $h$. The action
of the BPZ conjugation on the modes of $\tilde \phi(z)$ satisfies
the following useful property
\begin{align}
U_r^{\dag -1} \text{bpz}(\tilde \phi_n)
U_r^{\dag}=\big(\frac{2}{r}\big)^{-n} \text{bpz}(\tilde \phi_n) \,
.
\end{align}

Correlation functions which involve only modes of the $\tilde
c(z)$ ghost can be expressed in terms of the contour integral
(\ref{f}) as follows
\begin{align}
\label{co1}\langle \tilde{c}_p \tilde{c}_q \tilde{c}_r \rangle &=
\mathcal{F}(p-2,q-2,r-2,1,0,1,0,1,0) \, ,
\\
\label{corre2}\langle \text{bpz}(\tilde{c}_p) \tilde{c}_q
\tilde{c}_r \rangle &=
\mathcal{F}(p-2,q-2,r-2,1,\frac{\pi}{2},1,0,1,0)  \, ,
\\
\label{corre3}\langle \text{bpz}(\tilde{c}_p)
\text{bpz}(\tilde{c}_q) \tilde{c}_r \rangle &=
\mathcal{F}(p-2,q-2,r-2,1,\frac{\pi}{2},1,\frac{\pi}{2},1,0) \, .
\end{align}

Correlators involving modes of the $\tilde c(z)$ ghost and
insertions of operators $\hat{\mathcal{L}}^{n}$,
$\hat{\mathcal{B}}$, can be obtained by using the basic
correlators (\ref{a2})-(\ref{a5}) and the definition of
$\hat{\mathcal{L}}^{n} \equiv (-2)^n n! \oint \frac{dz}{2 \pi i}
\frac{1}{(z-2)^{n+1}} U^\dag_z U_z$. For instance, let us write a
correlator involving a $\hat{\mathcal{L}}^{n}$ insertion,
\begin{align}
\label{corre5}\langle
\text{bpz}(\tilde{c}_p)(\mathcal{L}_0+\mathcal{L}^\dag_0)^{n}
\tilde{c}_q \tilde{c}_r \rangle =(-1)^n n! {p+q+r \choose n}
\mathcal{F}(p-2,q-2,r-2,1,\frac{\pi}{2},1,0,1,0) \, .
\end{align}

As it was shown in reference \cite{AldoArroyo:2009hf}, the
computations of correlators involving the star $*$-product of
states defined in the sliver frame are straightforward. For
instance, let us compute the correlator $\langle
0|\text{bpz}(\tilde c_{p_1}) \hat{\mathcal{L}}^{n_1}
,\hat{\mathcal{L}}^{n_2}  \tilde c_{p_2} |0 \rangle *
\hat{\mathcal{L}}^{n_3}  \tilde c_{p_3}|0 \rangle$,
\begin{align}
&\label{corre7} \langle 0|\text{bpz}(\tilde c_{p_1})
\hat{\mathcal{L}}^{n_1} ,\hat{\mathcal{L}}^{n_2}  \tilde c_{p_2}
|0 \rangle * \hat{\mathcal{L}}^{n_3}  \tilde c_{p_3}|0 \rangle =
\nonumber \\
&= \frac{(-2)^{n_2+n_3} n_2! n_3! }{(2 \pi i)^{4}} \oint
\frac{dz_2 dz_3 dx_2 dx_3 \,
x_2^{p_2-2}x_3^{p_3-2}}{(z_2-2)^{n_2+1}(z_3-2)^{n_3+1}} \langle
0|\text{bpz}(\tilde c_{p_1}) \hat{\mathcal{L}}^{n_1}, U^\dag_{z_2}
U_{z_2} \tilde c(x_2)|0 \rangle * U^\dag_{z_3} U_{z_3} \tilde
c(x_3) |0 \rangle
\nonumber \\
&= \frac{(-2)^{n_2+n_3} n_2! n_3! }{(2 \pi i)^{4}} \oint
\frac{dz_2 dz_3 dx_2 dx_3 \,
x_2^{p_2-2}x_3^{p_3-2}}{(z_2-2)^{n_2+1}(z_3-2)^{n_3+1}} \times
\nonumber
\\ &\times \langle \text{bpz}(\tilde c_{p_1}) \hat{\mathcal{L}}^{n_1}
U^\dag_{r} U_{r} \tilde c(x_2+\frac{\pi}{4}(z_3-1))
\tilde c(x_3-\frac{\pi}{4}(z_2-1)) \rangle \nonumber \\
&=\frac{(-1)^{n_1+n_2+n_3} 2^{2 n_1+n_2+n_3-2p_1-4} n_1!n_2! n_3!
}{(2 \pi i)^{3}} \oint \frac{dz_1dz_2 dz_3 \,
z_1^{p_1+2}r^{p_1+2-n_1}}{(z_1-2)^{n_1+1}(z_2-2)^{n_2+1}(z_3-2)^{n_3+1}}
\times \nonumber
\\ &\times \mathcal{F}(p_1-2,p_2-2,p_3-2,1,\frac{\pi}{2},\frac{4}{z_1r},\frac{\pi(z_3-1)}{z_1r},
\frac{4}{z_1r},\frac{\pi(1-z_2)}{z_1r}) \, ,
\end{align}
where we have denoted $r\equiv z_2+z_3-1$, and the function
$\mathcal{F}$ was defined in equation (\ref{f}).

We have noted that all correlators involving the two point vertex,
namely the BPZ inner product, can be expressed in terms of the
function (\ref{f}) evaluated at some particular points without
performing any extra contour integral, e.g. equation
(\ref{corre5}). On the other hand correlators involving the three
point interaction vertex, namely correlators involving the
$*$-product can be expressed in terms of triple contour integrals
having as an integrand the function (\ref{f}), e.g. equation
(\ref{corre7}). In reference \cite{AldoArroyo:2009hf}, we dare to
compute explicitly these triple contour integrals, but without any
succeed, so we have written a computer program to do it for us.
Nevertheless to evaluate the tachyon potential using a string
field expanded in the $\mathcal{L}_0$-basis there is a technical
computational issue related to the evaluation of these triple
contour integrals \cite{AldoArroyo:2009hf}. Since the number of
states involved in the $\mathcal{L}_0$ level expansion of the
string field $\Psi$ grows exponentially with the level, the
complexity of performing many triple contour integrals, involved
in the evaluation of the three point interaction vertex $\langle
\Psi, \Psi*\Psi \rangle$, increases exponentially as well. To
solve this technical issue, it would be nice to find an
alternative and efficient method to evaluate correlators involving
the $*$-product.

Let us also note that we have just derived a straightforward way
of computing three point correlation functions involving operators
of the form $\hat{\mathcal{L}}^n \tilde c_p |0 \rangle$ and
$\hat{\mathcal{L}}^n \hat{\mathcal{B}} \tilde c_p \tilde c_q|0
\rangle$. However we can have more general set of states, like the
ones discussed in Schnabl's original paper \cite{Schnabl:2005gv},
for instance states of the form
\begin{align}
\label{ba01} \hat{\mathcal{L}}^m \mathcal{L}_{n_1} \cdots
\mathcal{L}_{n_k} \tilde c_p |0 \rangle \; &, \;\;\; \text{with}
\;\;\; n_1,\cdots,n_k =-2,-3,\cdots \\
\label{ba02} \hat{\mathcal{L}}^m \hat{\mathcal{B}}
\mathcal{L}_{n_1} \cdots \mathcal{L}_{n_k} \tilde c_p \tilde c_q
|0 \rangle \; &, \;\;\; \text{with} \;\;\; n_1,\cdots,n_k
=-2,-3,\cdots \\
\label{ba03} \hat{\mathcal{L}}^m \mathcal{L}_{n_1} \cdots
\mathcal{L}_{n_k} \mathcal{B}_{l} \tilde c_p \tilde c_q |0 \rangle
\; &, \;\;\; \text{with}
\;\;\; n_1,\cdots,n_k,l =-2,-3,\cdots \\
\label{ba04} \hat{\mathcal{L}}^m \hat{\mathcal{B}}
\mathcal{L}_{n_1} \cdots \mathcal{L}_{n_k}  \mathcal{B}_{l} \tilde
c_p \tilde c_q \tilde c_r |0 \rangle \; & , \;\;\; \text{with}
\;\;\; n_1,\cdots,n_k,l =-2,-3,\cdots \;
\end{align}
where the operators $\mathcal{L}_{n_k}$ and $\mathcal{B}_{l}$ are
modes of the worldsheet energy momentum tensor and the $b$ ghost
in the $\tilde z$ coordinate (sliver frame). By a conformal
transformation they can be written as
\begin{eqnarray}
\label{llm} \mathcal{L}_{n_k} &=& \oint \frac{d z}{2 \pi i}
(1+z^2)
(\arctan z)^{n_k+1} \, T(z) \, ,\\
\label{llm} \mathcal{B}_{l} &=& \oint \frac{d z}{2 \pi i} (1+z^2)
(\arctan z)^{l+1} \, b(z) \, .
\end{eqnarray}

Based on conservation laws \cite{Rastelli:2000iu}, we will derive
an alternative method for computing the three point interaction
vertex defined in the sliver frame. The advantage of this
alternative method is that the vertex operators involved in the
evaluation of the three point correlation function are generic
ones, not just the operators discussed in reference
\cite{AldoArroyo:2009hf} but also operators like the ones given in
(\ref{ba01})-(\ref{ba04}). Schematically, in this alternative
method based on conservation laws, the task consists in finding
(inside the three point correlation function) an expression for
the operators $\hat{\mathcal{L}}^m$, $\hat{\mathcal{B}}$,
$\mathcal{L}_{n_k}$ and $\mathcal{B}_{l}$ (with $n_k,l
=-2,-3,\cdots$) in terms of the modes $\mathcal{L}_{u}$ and
$\mathcal{B}_{v}$ with $u,v =-1,0,1,\cdots$. Since for these modes
we have $\mathcal{L}_{u} |0 \rangle=0$, $\mathcal{B}_{v} |0
\rangle=0$, the commutation and anticommutation relations of
$\mathcal{L}_{u}$ and $\mathcal{B}_v$ with the rest of operators
allow to find a recursive procedure that express the three point
correlation function of any state in terms of correlators which
only involve the modes of the $c$ ghost.

\section{Conservation laws}
In this section, we will discuss in detail the derivation of
conservation laws for operators which are widely used in the
$\mathcal{L}_0$ level expansion of the string field. We begin by
reviewing the derivation of the conservation laws of a rather
generic conformal field and then apply the result to the case of
operators defined on the sliver frame.

Suppose we have a conformal field $\phi$ of weight $h$. Let
$\upsilon(z)$ be a field of conformal weight $1-h$, then the
quantity $\upsilon (z) \phi(z) dz$ transform as a 1-form. We
require $\upsilon(z)$ to be holomorphic everywhere in the $z$
plane, except at the punctures where it may have poles. For this
1-form $\upsilon (z) \phi(z) dz$ to be regular at infinity,
$\lim_{z \rightarrow \infty} z^{2-2h} \upsilon(z)$ must be
constant (or zero). Thanks to the holomorphicity of $\upsilon(z)$,
integration contours in the $z$ plane can be continuously deformed
as long as we do not cross a puncture. Consider a contour
$\mathcal{C}$ which encircles the $n$ punctures at $f^n_1(0),
\cdots ,f^n_n(0)$. For arbitrary vertex operators $\Phi_i$, the
correlator
\begin{eqnarray}
\label{corre1section3} \big\langle \oint_{\mathcal{C}} \frac{1}{2
\pi i} \upsilon(z) \phi(z) dz f^n_1 \circ \Phi_1(0) \cdots f^n_n
\circ \Phi_n(0)
 \big\rangle
\end{eqnarray}
vanishes identically, by shrinking the contour $\mathcal{C}$ to
zero size around the point at infinity.

Since the $n$-point correlation function (\ref{corre1section3}) is
zero for arbitrary vertex operators $\Phi_i$, we can write
\begin{eqnarray}
\label{conserva1section3} \big\langle V_n \big|
\oint_{\mathcal{C}} \frac{1}{2 \pi i} \upsilon(z) \phi(z) dz = 0.
\end{eqnarray}
Deforming the contour $\mathcal{C}$ into the sum of $n$ contours
$\mathcal{C}_i$ around the $n$ punctures, and referring the 1-form
to the local coordinates, we obtain the basic relation
\begin{eqnarray}
\label{conserva2section3} \big\langle V_n \big| \sum_{i=1}^{n}
\oint_{\mathcal{C}_i} \frac{1}{2 \pi i} \upsilon^{(i)}(z_i)
\phi(z_i) dz_i = 0 \, ,
\end{eqnarray}
where $\upsilon ^{(i)}(z_i) = ( \partial_{z_i}f^{n}_{i}
(z_i))^{1-h} \upsilon(f^n_i (z_i)) $. The notation $\big\langle
V_n \big|$ represents the number of interaction vertices
\cite{Rastelli:2000iu}.

\subsection{Virasoro conservation laws}
Let us recall that the operators used in the $\mathcal{L}_0$-basis
are given in terms of the basic operators $\hat{\mathcal{L}}$,
$\hat{\mathcal{B}}$ and $\tilde c_p$. These operators are related
to the worldsheet energy momentum tensor $T(z)$, the $b(z)$ and
$c(z)$ ghosts fields respectively. In this subsection we will
derive the conservation law for the $\hat{\mathcal{L}}$ operator
\begin{eqnarray}
\label{lhat1} \hat{\mathcal{L}}= \oint \frac{d z}{2 \pi i}
(1+z^{2}) (\arctan z+\text{arccot} z) \, T(z) \, .
\end{eqnarray}

Since we are interested in the conservation laws of operators
defined on the sliver frame. Using the conformal mapping $\tilde
z= \arctan z $, we can write the expression of the
$\hat{\mathcal{L}}$ operator as it is given in the sliver frame
\begin{eqnarray}
\label{lhat2} \hat{\mathcal{L}}=\frac{\pi}{2} \oint \frac{d \tilde
z}{2 \pi i} \varepsilon (\text{Re} \tilde z) \, \tilde T( \tilde
z) \, ,
\end{eqnarray}
where $\varepsilon (x)$ is the step function equal to $\pm 1$ for
positive or negative values of its argument respectively.

For vertex operators $\Phi_i$ defined on the sliver frame, the
three functions $f_1$, $f_2$ and $f_3$ which appear in the
definition of the three point vertex $\big\langle f_1 \circ
\Phi_1(0) f_2 \circ \Phi_2(0) f_3 \circ \Phi_3(0)
 \big\rangle$ are given by
\begin{eqnarray}
\label{fff1} f_1(\tilde z_1) &=& \tan (\frac{\pi}{3} + \frac{2}{3}\tilde z_1) \, , \\
\label{fff2} f_2(\tilde z_2) &=& \tan ( \frac{2}{3}\tilde z_2) \, , \\
\label{fff3} f_3(\tilde z_3) &=& \tan (-\frac{\pi}{3} +
\frac{2}{3}\tilde z_3) \, .
\end{eqnarray}

We are looking for conservation laws such that the operator
$\hat{\mathcal{L}}$ acting on the three point interaction vertex
$\big\langle V_3 \big|$ can be expressed in terms of positive
Virasoro modes defined on the sliver frame\footnote{We are going
to use the following notation $\mathcal{O}^{(i)}$ to refer an
operator $\mathcal{O}$ defined around the $i$-th puncture.}
\begin{eqnarray}
\label{V3Lhat1} \big\langle V_3 \big| \hat{\mathcal{L}}^{(2)} =
\big\langle V_3 \big| \Big(  \sum_{n = 0}^{\infty}  a_n
\mathcal{L}_n^{(1)} + \sum_{n = 0}^{\infty}  b_n
\mathcal{L}_n^{(2)}
 + \sum_{n = 0}^{\infty}  c_n
\mathcal{L}_n^{(3)}  \Big) \, ,
\end{eqnarray}
where $a_n$, $b_n$ and $c_n$ are coefficients that will be
determined below and depend on the geometry of the vertex. By the
cyclicity property of the three vertex, the same identity
(\ref{V3Lhat1}) holds after letting (1) $\rightarrow$ (2), (2)
$\rightarrow$ (3), (3) $\rightarrow$ (1).

In order to find conservation laws of the form (\ref{V3Lhat1}), we
need a vector field which behaves as $ v^{(2)}(\tilde z_2) \sim
\frac{\pi}{2} \varepsilon (\text{Re} \tilde z_2) + O(\tilde z_2) $
around puncture 2, and has the following behavior in the other two
punctures, $v^{(1)}(\tilde z_1) \sim O(\tilde z_3)$,
$v^{(3)}(\tilde z_3) \sim O(\tilde z_3) $. A vector field which
does this job is given by
\begin{eqnarray}
\label{vv1} v(z) = \frac{2}{3} (1+z^2) \text{arccot} z - \frac{4
\pi}{9 \sqrt 3} z \, .
\end{eqnarray}
This vector field is not globally defined, but is holomorphic
everywhere outside the unit circle including the infinity, so that
(\ref{corre1section3}) still holds for a contour encircling the
infinity.

Using equations (\ref{fff1})-(\ref{fff3}) and (\ref{vv1}) into the
definition $v^{(i)}(\tilde z_i) = (
\partial_{\tilde{z}_i}f_{i} (\tilde{z}_i))^{-1} v(f_i (\tilde z_i))$
of the vector fields $v^{(1)}(\tilde z_1)$, $v^{(2)}(\tilde z_2)$
and $v^{(3)}(\tilde z_3)$, we find that
\begin{eqnarray}
v^{(1)}(\tilde z_1) &=& \big(-\frac{2}{3}+\frac{2 \pi }{9
\sqrt{3}}\big) \tilde z_1+\frac{4 \pi  \tilde z_1^2}{27}-\frac{16
\pi
   \tilde z_1^3}{243 \sqrt{3}}-\frac{16 \pi  \tilde z_1^4}{729}+\frac{64 \pi  \tilde z_1^5}{10935
   \sqrt{3}} + O(\tilde z_1^6)\, , \;\;\;\;\;\; \\
v^{(2)}(\tilde z_2) &=& \frac{\pi}{2} \varepsilon (\text{Re}
\tilde z_2)+\big(-\frac{2}{3}-\frac{4 \pi }{9 \sqrt{3}}\big)
\tilde z_2+\frac{32 \pi  \tilde z_2^3}{243 \sqrt{3}}-\frac{128
   \pi  \tilde z_2^5}{10935 \sqrt{3}} + O(\tilde z_2^7)\, , \;\;\;\;\;\; \\
v^{(3)}(\tilde z_3) &=& \big(-\frac{2}{3}+\frac{2 \pi }{9
\sqrt{3}}\big) \tilde z_3-\frac{4 \pi  \tilde z_3^2}{27}-\frac{16
\pi
   \tilde z_3^3}{243 \sqrt{3}}+\frac{16 \pi  \tilde z_3^4}{729}+\frac{64 \pi  \tilde z_3^5}{10935
   \sqrt{3}} + O(\tilde z_3^6) \, . \;\;\;\;\;\;
\end{eqnarray}
Due to the presence of the step function we see that the vector
field $v^{(2)}(\tilde z_2)$ is discontinuous around puncture 2,
since we are interesting in the conservation law of the operator
defined in equation (\ref{lhat2}), this kind of discontinuity is
what we want. Using (\ref{conserva2section3}) and noting that
integration amounts to the replacement $v^{(i)}_n \tilde z_i^n
\rightarrow v^{(i)}_n \mathcal{L}^{(i)}_ {n-1}$, we can
immediately write the conservation law
\begin{align}
0=&\big\langle V_3 \big| \Big( \big(-\frac{2}{3}+\frac{2 \pi }{9
\sqrt{3}}\big) \mathcal{L}_0+\frac{4 \pi
}{27}\mathcal{L}_1-\frac{16 \pi}{243
\sqrt{3}}\mathcal{L}_2-\frac{16 \pi}{729}\mathcal{L}_3+\frac{64
\pi }{10935
   \sqrt{3}} \mathcal{L}_4 +\cdots \Big)^{(1)} + \nonumber \\
&+\big\langle V_3 \big| \Big(
\hat{\mathcal{L}}+\big(-\frac{2}{3}-\frac{4 \pi }{9 \sqrt{3}}\big)
\mathcal{L}_0+\frac{32 \pi }{243 \sqrt{3}}\mathcal{L}_2-\frac{128
   \pi }{10935 \sqrt{3}}\mathcal{L}_4 +\cdots \Big)^{(2)} +
   \nonumber \\
\label{conserv01} &+\big\langle V_3 \big| \Big(
\big(-\frac{2}{3}+\frac{2 \pi }{9 \sqrt{3}}\big)
\mathcal{L}_0-\frac{4 \pi }{27}\mathcal{L}_1-\frac{16 \pi}{243
\sqrt{3}}\mathcal{L}_2+\frac{16 \pi}{729}\mathcal{L}_3+\frac{64
\pi }{10935
   \sqrt{3}} \mathcal{L}_4 +\cdots \Big)^{(3)} \, .
\end{align}
Thanks to the cyclicity of the string vertex, analogous identities
hold by cycling the punctures, (1) $\rightarrow$ (2), (2)
$\rightarrow$ (3), (3) $\rightarrow$ (1).

Let us recall that the $\mathcal{L}_0$-basis contains operators of
the form $\hat{\mathcal{L}}^N$ with $N=1,2,\cdots$ At this point
we just have shown the conservation law for the
$\hat{\mathcal{L}}$ operator (\ref{conserv01}), namely the $N=1$
case. It remains to find conservation laws for the other values of
$N$, for instance the $N=2$ case reads
\begin{eqnarray}
 \big\langle V_3 \big| \hat{\mathcal{L}}^{(2)}
\hat{\mathcal{L}}^{(2)} &=& \big\langle V_3 \big| \Big(  \sum_{n =
0}^{\infty}  a_n \mathcal{L}_n^{(1)} + \sum_{n = 0}^{\infty}  b_n
\mathcal{L}_n^{(2)}
 + \sum_{n = 0}^{\infty}  c_n
\mathcal{L}_n^{(3)}  \Big) \hat{\mathcal{L}}^{(2)} \nonumber \\
&=& \big\langle V_3 \big| \Big(  \sum_{n = 0}^{\infty}  a_n
\mathcal{L}_n^{(1)} + \sum_{n = 0}^{\infty}  b_n
\mathcal{L}_n^{(2)}
 + \sum_{n = 0}^{\infty}  c_n
\mathcal{L}_n^{(3)}  \Big)^2 + \nonumber \\ \label{V3Lhat2} &&
\big\langle V_3 \big| \Big[ \Big( \sum_{n = 0}^{\infty}  a_n
\mathcal{L}_n^{(1)} + \sum_{n = 0}^{\infty} b_n
\mathcal{L}_n^{(2)}
 + \sum_{n = 0}^{\infty}  c_n
\mathcal{L}_n^{(3)}  \Big), \hat{\mathcal{L}}^{(2)} \Big] .
\end{eqnarray}
From equation (\ref{V3Lhat2}) it is clear that for each $n \geq
0$, we will have to compute additional conservation laws of the
form
\begin{eqnarray}
\label{vv33} \big\langle V_3 \big| [\mathcal{L}_n^{(i)},
\hat{\mathcal{L}}^{(2)}] = \big\langle V_3 \big| \Big(
 \sum_{k = 0}^{\infty}  \alpha^{(i)}_{n,k} \mathcal{L}_k^{(1)} + \sum_{k = 0}^{\infty}
\beta^{(i)}_{n,k} \mathcal{L}_k^{(2)}
 + \sum_{k = 0}^{\infty}  \gamma^{(i)}_{n,k}
\mathcal{L}_k^{(3)}  \Big) .
\end{eqnarray}
Plugging this equation (\ref{vv33}) into the last term of equation
(\ref{V3Lhat2}) produces double sums (running over the indices $n$
and $k$). Following the same procedure as for the $N=2$ case, for
the $N=3$ case we will get triple sums, for the $N=4$ case
four-fold sums and so on and so forth. Therefore due to the
presence of these $N$-fold sums, it seems that this alternative
method based on conservation laws should not be efficient for
computing three point correlation functions involving operators of
the form $\hat{\mathcal{L}}^N$. Nevertheless, as a remarkably
result, in what follows, we will show that these $N$-fold sums can
be avoided.

The idea is to find a conservation law which contains almost all
the terms given in equation (\ref{conserv01}), except for some
finite terms. Using equations (\ref{fff1})-(\ref{fff3}) and the
vector field $\hat{v}(z)= 4 \pi z/(9 \sqrt{3})  $ into the
definition $\hat{v} ^{(i)}(\tilde z_i) = (
\partial_{\tilde{z}_i}f_{i} (\tilde{z}_i))^{-1} \hat{v}(f_i (\tilde z_i))$
of the vector fields $\hat{v}^{(1)}(\tilde z_1)$,
$\hat{v}^{(2)}(\tilde z_2)$ and $\hat{v}^{(3)}(\tilde z_3)$, we
find that
\begin{eqnarray}
\hat{v}^{(1)}(\tilde z_1) &=& \frac{\pi }{6}-\frac{2 \pi \tilde
z_1}{9 \sqrt{3}}-\frac{4 \pi  \tilde z_1^2}{27}+\frac{16 \pi
\tilde z_1^3}{243 \sqrt{3}}+\frac{16
   \pi  \tilde z_1^4}{729}-\frac{64 \pi  \tilde z_1^5}{10935 \sqrt{3}}+O(\tilde z_1^6)\, , \;\;\;\;\;\; \\
\hat{v}^{(2)}(\tilde z_2) &=& \frac{4 \pi  \tilde z_2}{9
\sqrt{3}}-\frac{32 \pi  \tilde z_2^3}{243 \sqrt{3}}+\frac{128 \pi
\tilde z_2^5}{10935
   \sqrt{3}} + O(\tilde z_2^7)\, , \;\;\;\;\;\; \\
\hat{v}^{(3)}(\tilde z_3) &=& -\frac{\pi }{6}-\frac{2 \pi \tilde
z_3}{9 \sqrt{3}}+\frac{4 \pi  \tilde z_3^2}{27}+\frac{16 \pi
\tilde z_3^3}{243
   \sqrt{3}}-\frac{16 \pi  \tilde z_3^4}{729}-\frac{64 \pi  \tilde z_3^5}{10935 \sqrt{3}} + O(\tilde z_3^6) \, . \;\;\;\;\;\;
\end{eqnarray}
Using (\ref{conserva2section3}) and noting that integration
amounts to the replacement $\hat{v}^{(i)}_n \tilde z_i^n
\rightarrow \hat{v}^{(i)}_n \mathcal{L}^{(i)}_ {n-1}$, we can
immediately write the conservation law
\begin{align}
0=&\big\langle V_3 \big| \Big( \frac{\pi}{6} \mathcal{L}_{-1}
-\frac{2 \pi }{9 \sqrt{3}} \mathcal{L}_0-\frac{4 \pi
}{27}\mathcal{L}_1+\frac{16 \pi}{243
\sqrt{3}}\mathcal{L}_2+\frac{16 \pi}{729}\mathcal{L}_3-\frac{64
\pi }{10935
   \sqrt{3}} \mathcal{L}_4 +\cdots \Big)^{(1)} + \nonumber \\
&+\big\langle V_3 \big| \Big( \frac{4 \pi }{9 \sqrt{3}}
\mathcal{L}_0-\frac{32 \pi }{243 \sqrt{3}}\mathcal{L}_2+\frac{128
   \pi }{10935 \sqrt{3}}\mathcal{L}_4 +\cdots \Big)^{(2)} +
   \nonumber \\
\label{conserv0002} &+\big\langle V_3 \big| \Big( -\frac{\pi}{6}
\mathcal{L}_{-1} -\frac{2 \pi }{9 \sqrt{3}} \mathcal{L}_0+\frac{4
\pi }{27}\mathcal{L}_1+\frac{16 \pi}{243
\sqrt{3}}\mathcal{L}_2-\frac{16 \pi}{729}\mathcal{L}_3-\frac{64
\pi }{10935
   \sqrt{3}} \mathcal{L}_4 +\cdots \Big)^{(3)} \, .
\end{align}
Adding the equations (\ref{conserv0002}) and (\ref{conserv01}), we
obtain the following conservation law for the $\hat{\mathcal{L}}$
operator
\begin{align}
\label{conserv0003} \big\langle V_3 \big| \hat{\mathcal{L}}^{(2)}
= \big\langle V_3 \big| \Big( \frac{2}{3} \big(
\mathcal{L}_0^{(1)} + \mathcal{L}_0^{(2)} + \mathcal{L}_0^{(3)}
\big) + \frac{\pi}{6} \mathcal{L}_{-1}^{(3)} -  \frac{\pi}{6}
\mathcal{L}_{-1}^{(1)}
 \Big).
\end{align}
Thanks to the cyclicity of the string vertex, analogous identities
hold by cycling the punctures, (1) $\rightarrow$ (2), (2)
$\rightarrow$ (3), (3) $\rightarrow$ (1)
\begin{align}
\label{conservxx1} \big\langle V_3 \big| \hat{\mathcal{L}}^{(1)}
&= \big\langle V_3 \big| \Big( \frac{2}{3} \big(
\mathcal{L}_0^{(1)} + \mathcal{L}_0^{(2)} + \mathcal{L}_0^{(3)}
\big) + \frac{\pi}{6} \mathcal{L}_{-1}^{(2)} -  \frac{\pi}{6}
\mathcal{L}_{-1}^{(3)}
 \Big), \\
\label{conservxx2} \big\langle V_3 \big| \hat{\mathcal{L}}^{(3)}
&= \big\langle V_3 \big| \Big( \frac{2}{3} \big(
\mathcal{L}_0^{(1)} + \mathcal{L}_0^{(2)} + \mathcal{L}_0^{(3)}
\big) + \frac{\pi}{6} \mathcal{L}_{-1}^{(1)} -  \frac{\pi}{6}
\mathcal{L}_{-1}^{(2)}
 \Big).
\end{align}

Since the conservation law (\ref{conserv0003}) involves a finite
numbers of terms containing the $\mathcal{L}_0^{(i)}$ and
$\mathcal{L}_{-1}^{(i)}$ operators, using the commutators
$[\mathcal{L}_0^{(i)},\hat{\mathcal{L}}^{(j)}]=\delta^{ij}\hat{\mathcal{L}}^{(j)}$
and $[\mathcal{L}_{-1}^{(i)},\hat{\mathcal{L}}^{(j)}]=0$, we can
easily derive conservation laws for the operators
$\hat{\mathcal{L}}^N$ with $N \geq 2$. For instance the $N=2$ case
reads
\begin{eqnarray}
 \big\langle V_3 \big| \hat{\mathcal{L}}^{(2)}
\hat{\mathcal{L}}^{(2)} &=& \big\langle V_3 \big| \Big(
\frac{2}{3} \big( \mathcal{L}_0^{(1)} + \mathcal{L}_0^{(2)} +
\mathcal{L}_0^{(3)} \big) + \frac{\pi}{6} \mathcal{L}_{-1}^{(3)} -
\frac{\pi}{6}
\mathcal{L}_{-1}^{(1)} \Big)^2 + \nonumber \\
\label{V3Lhat0005} && \big\langle V_3 \big|  \Big( \frac{4}{9}
\big( \mathcal{L}_0^{(1)} + \mathcal{L}_0^{(2)} +
\mathcal{L}_0^{(3)} \big) + \frac{\pi}{9} \mathcal{L}_{-1}^{(3)} -
\frac{\pi}{9} \mathcal{L}_{-1}^{(1)} \Big) .
\end{eqnarray}
Again due to the cyclicity of the string vertex, analogous
identities hold by cycling the punctures, (1) $\rightarrow$ (2),
(2) $\rightarrow$ (3), (3) $\rightarrow$ (1). Given the simplicity
of the conservation law (\ref{V3Lhat0005}) compared to the one
given in equation (\ref{V3Lhat2}), we observe that, based on
conservation laws, we can get an efficient and alternative method
for computing three point correlation functions involving
operators of the form $\hat{\mathcal{L}}^N$.

For completeness reasons, we also write conservation laws for some
negative modes $\mathcal{L}_{-k}$ with $k=2,3,\cdots$ of the
Virasoro operators defined on the sliver frame.  With
\begin{eqnarray}
\label{phi2cL} v_2(z) &=&-\frac{4 \left(z^2-3\right)}{27 z}
\, , \\
\label{phi3cL} v_3(z) &=&-\frac{8 \left(z^2-3\right)}{81 z^2}-\frac{16}{81} \left(z^2-3\right) \, , \\
\label{phi4cL} v_4(z) &=& -\frac{16 \left(z^2-3\right)}{243
z^3}+\frac{28 }{27}v_2(z) \, ,
\end{eqnarray}
we obtain the conservation laws
\begin{align}
0=&\big\langle V_3 \big| \Big(-\frac{8}{27}
\mathcal{L}_0+\frac{80}{81\sqrt{3}} \mathcal{L}_1-\frac{64}{243}
\mathcal{L}_2+\frac{64}{729\sqrt{3}} \mathcal{L}_3-\frac{5888}{98415} \mathcal{L}_4+\frac{512}{6561\sqrt{3}} \mathcal{L}_5+\cdots \Big)^{(1)} + \nonumber \\
&+\big\langle V_3 \big| \Big( \mathcal{L}_{-2} -\frac{20 }{27}
\mathcal{L}_0 +\frac{208}{1215} \mathcal{L}_2 -\frac{2176}{137781}
\mathcal{L}_4 +\cdots \Big)^{(2)} +
   \nonumber \\
\label{conservcc02Lx2} &+\big\langle V_3 \big| \Big( -\frac{8}{27}
\mathcal{L}_0-\frac{80}{81\sqrt{3}} \mathcal{L}_1-\frac{64}{243}
\mathcal{L}_2-\frac{64}{729\sqrt{3}}
\mathcal{L}_3-\frac{5888}{98415}
\mathcal{L}_4-\frac{512}{6561\sqrt{3}} \mathcal{L}_5+\cdots
\Big)^{(3)}
   . \\
0=&\big\langle V_3 \big| \Big(-\frac{112 }{81 \sqrt{3}}
\mathcal{L}_0+\frac{160 }{243}\mathcal{L}_1-\frac{128
 }{243 \sqrt{3}} \mathcal{L}_2+\frac{6272 }{19683}\mathcal{L}_3-\frac{14848
  }{32805 \sqrt{3}}\mathcal{L}_4+\frac{11264
   }{59049}\mathcal{L}_5 +\cdots \Big)^{(1)} + \nonumber \\
&+\big\langle V_3 \big| \Big( \mathcal{L}_{-3}  -\frac{304 }{1215}
\mathcal{L}_1+\frac{5504
   }{137781} \mathcal{L}_3-\frac{3328
   }{1476225} \mathcal{L}_5+\cdots
\Big)^{(2)} +
   \nonumber \\
\label{conservcc02Lx3} &+\big\langle V_3 \big| \Big( \frac{112
}{81 \sqrt{3}} \mathcal{L}_0+\frac{160
}{243}\mathcal{L}_1+\frac{128
 }{243 \sqrt{3}} \mathcal{L}_2+\frac{6272 }{19683}\mathcal{L}_3+\frac{14848
  }{32805 \sqrt{3}}\mathcal{L}_4+\frac{11264
   }{59049}\mathcal{L}_5 +\cdots \Big)^{(3)}
   .\\
 0=&\big\langle V_3 \big| \Big( -\frac{256 }{729} \mathcal{L}_0+\frac{1024 }{729
   \sqrt{3}}\mathcal{L}_1-\frac{14336 }{19683}\mathcal{L}_2+\frac{20480
   }{19683 \sqrt{3}}\mathcal{L}_3-\frac{499712
   }{885735}\mathcal{L}_4+\frac{6651904 }{7971615
   \sqrt{3}}\mathcal{L}_5 +\cdots \Big)^{(1)} + \nonumber \\
&+\big\langle V_3 \big| \Big( \mathcal{L}_{-4} -\frac{1328
}{3645}\mathcal{L}_0+\frac{14080
   }{137781} \mathcal{L}_2-\frac{41728
  }{4428675} \mathcal{L}_4+\cdots
\Big)^{(2)} +
   \nonumber \\
\label{conservcc02Lx4} &+\big\langle V_3 \big| \Big( -\frac{256
}{729} \mathcal{L}_0-\frac{1024 }{729
   \sqrt{3}}\mathcal{L}_1-\frac{14336 }{19683}\mathcal{L}_2-\frac{20480
   }{19683 \sqrt{3}}\mathcal{L}_3-\frac{499712
   }{885735}\mathcal{L}_4-\frac{6651904 }{7971615
   \sqrt{3}}\mathcal{L}_5+\cdots
\Big)^{(3)}.
\end{align}
Thanks to the cyclicity of the string vertex, analogous identities
hold by cycling the punctures, (1) $\rightarrow$ (2), (2)
$\rightarrow$ (3), (3) $\rightarrow$ (1).

\subsection{Ghost conservation laws}
Since the $b$ ghost is a conformal field of dimension two, the
conservation laws for operators involving this field are identical
to those for the Virasoro operators. For instance, the
conservation laws for the operator $\hat{\mathcal{B}}$ reads
\begin{align}
\label{conserv00007} \big\langle V_3 \big| \hat{\mathcal{B}}^{(1)}
&= \big\langle V_3 \big| \Big( \frac{2}{3} \big(
\mathcal{B}_0^{(1)} + \mathcal{B}_0^{(2)} + \mathcal{B}_0^{(3)}
\big) + \frac{\pi}{6} \mathcal{B}_{-1}^{(2)} -  \frac{\pi}{6}
\mathcal{B}_{-1}^{(3)}
 \Big), \\
\big\langle V_3 \big| \hat{\mathcal{B}}^{(2)} &= \big\langle V_3
\big| \Big( \frac{2}{3} \big( \mathcal{B}_0^{(1)} +
\mathcal{B}_0^{(2)} + \mathcal{B}_0^{(3)} \big) + \frac{\pi}{6}
\mathcal{B}_{-1}^{(3)} -  \frac{\pi}{6} \mathcal{B}_{-1}^{(1)}
 \Big), \\
 \big\langle V_3 \big| \hat{\mathcal{B}}^{(3)}
&= \big\langle V_3 \big| \Big( \frac{2}{3} \big(
\mathcal{B}_0^{(1)} + \mathcal{B}_0^{(2)} + \mathcal{B}_0^{(3)}
\big) + \frac{\pi}{6} \mathcal{B}_{-1}^{(1)} -  \frac{\pi}{6}
\mathcal{B}_{-1}^{(2)}
 \Big).
\end{align}

Therefore it remains the derivation of conservation laws for the
$c$ ghost. The $c$ ghost is a primary field of dimension minus
one, thus to derive its conservation laws we need to consider a
globally defined quadratic differential
\begin{eqnarray}
 \varphi(z) (dz)^2 = \varphi(z') (dz')^2 ,
\end{eqnarray}
holomorphic everywhere except for possible poles at the punctures.
Regularity at infinity requires the $\lim_{z \rightarrow \infty}
z^4 \varphi(z)$ to be finite. The product $c(z)\varphi(z)dz$ is a
1- form, we can then use contour deformations and following
exactly the same procedure as for the 1-form considered in the
previous subsection, we derive
\begin{eqnarray}
\label{conservaghost1} \big\langle V_3 \big| \sum_{i=1}^{3}
\oint_{\mathcal{C}_i} \frac{1}{2 \pi i} \varphi^{(i)}(\tilde z_i)
\tilde c(\tilde z_i) d\tilde z_i = 0 \, ,
\end{eqnarray}
where $\varphi^{(i)}(\tilde z_i) = ( \partial_{\tilde z_i}f_{i}
(\tilde z_i))^{2} \varphi(f_i (\tilde z_i)) $, and since we are
considering conservation laws for operators defined on the sliver
frame, the functions $f_1$, $f_2$ and $f_3$ are the same as the
ones given in equations (\ref{fff1})-(\ref{fff3}).

For instance, with the quadratic differential
\begin{eqnarray}
\label{phi0c} \varphi_0(z) = -\frac{3}{z^2 \left(z^2-3\right)} \,
,
\end{eqnarray}
using the definition $\varphi_0^{(i)}(\tilde z_i) = (
\partial_{\tilde z_i}f_{i} (\tilde z_i))^{2} \varphi_0(f_i (\tilde
z_i)) $ from equation (\ref{conservaghost1}), we derive the
conservation law
\begin{align}
0=&\big\langle V_3 \big| \Big( -\frac{4}{3 \sqrt{3}} \tilde
c_1+\frac{8}{27}\tilde c_2-\frac{112 }{81 \sqrt{3}} \tilde
c_3+\frac{256 }{729} \tilde c_4-\frac{3392
   }{3645 \sqrt{3}} \tilde c_5 +\frac{5120 }{19683} \tilde c_6 +\cdots \Big)^{(1)} + \nonumber \\
&+\big\langle V_3 \big| \Big( \tilde c_0 +\frac{20}{27}\tilde
c_2+\frac{1328}{3645} \tilde c_4+\frac{21632 }{137781} \tilde c_6
+\cdots \Big)^{(2)} +
   \nonumber \\
\label{conservcc01} &+\big\langle V_3 \big| \Big( \frac{4}{3
\sqrt{3} } \tilde c_1+\frac{8}{27} \tilde c_2+\frac{112 }{81
\sqrt{3}} \tilde c_3 +\frac{256 }{729} \tilde c_4+\frac{3392
}{3645
   \sqrt{3}} \tilde c_5+\frac{5120 }{19683} \tilde c_6+\cdots \Big)^{(3)} \, .
\end{align}

Conservation laws for higher negative modes of the $c$ ghost are
obtained with quadratic differentials having higher order poles at
$z = 0$. With
\begin{eqnarray}
\label{phi1c} \varphi_1(z) &=& -\frac{2}{z^3 \left(z^2-3\right)}
\, , \\
\label{phi2c} \varphi_2(z) &=& -\frac{4}{3 z^4
\left(z^2-3\right)}-\frac{4}{9} \varphi _0(z) \, , \\
\label{phi3c} \varphi_3(z) &=& -\frac{8}{9 z^5
\left(z^2-3\right)}-\frac{8 }{27}\varphi_1(z) \, ,
\end{eqnarray}
we obtain the conservation laws
\begin{align}
0=&\big\langle V_3 \big| \Big(-\frac{8}{27} \tilde
c_1+\frac{80}{81 \sqrt{3}} \tilde c_2-\frac{160 }{243} \tilde
c_3+\frac{1024 }{729 \sqrt{3}} \tilde c_4-\frac{71552
   }{98415} \tilde c_5+\frac{8192 }{6561 \sqrt{3}} \tilde c_6+\cdots \Big)^{(1)} + \nonumber \\
&+\big\langle V_3 \big| \Big( \tilde c_{-1}+\frac{16}{27} \tilde
c_1+\frac{304 }{1215} \tilde c_3+\frac{68608
   }{688905} \tilde c_5+\cdots \Big)^{(2)} +
   \nonumber \\
\label{conservcc02} &+\big\langle V_3 \big| \Big( -\frac{8}{27 }
\tilde c_1-\frac{80}{81 \sqrt{3}} \tilde c_2-\frac{160 }{243}
\tilde c_3-\frac{1024 }{729 \sqrt{3}} \tilde c_4-\frac{71552
   }{98415} \tilde c_5-\frac{8192 }{6561 \sqrt{3}} \tilde c_6+\cdots \Big)^{(3)}
   . \\
0=&\big\langle V_3 \big| \Big(\frac{32}{81 \sqrt{3} } \tilde
c_1+\frac{64}{243} \tilde c_2-\frac{128 }{243 \sqrt{3}} \tilde
c_3+\frac{14336 }{19683} \tilde c_4-\frac{45568
   }{32805 \sqrt{3}} \tilde c_5+\frac{57344 }{59049} \tilde c_6 +\cdots \Big)^{(1)} + \nonumber \\
&+\big\langle V_3 \big| \Big( \tilde c_{-2}-\frac{208}{1215}
\tilde c_2-\frac{14080 }{137781} \tilde c_4-\frac{13568
   }{295245} \tilde c_6
+\cdots \Big)^{(2)} +
   \nonumber \\
\label{conservcc03} &+\big\langle V_3 \big| \Big( -\frac{32}{81
\sqrt{3} } \tilde c_1+\frac{64}{243} \tilde c_2+\frac{128 }{243
\sqrt{3}} \tilde c_3+\frac{14336
   }{19683} \tilde c_4+\frac{45568 }{32805 \sqrt{3}} \tilde c_5+\frac{57344 }{59049} \tilde c_6+\cdots \Big)^{(3)}
   . \\
 0=&\big\langle V_3 \big| \Big( \frac{32}{729 } \tilde c_1+\frac{64}{729 \sqrt{3}} \tilde c_2-\frac{6272 }{19683} \tilde c_3+\frac{20480 }{19683
   \sqrt{3}} \tilde c_4-\frac{833024 }{885735} \tilde c_5+\frac{3276800 }{1594323 \sqrt{3}} \tilde c_6 +\cdots \Big)^{(1)} + \nonumber \\
&+\big\langle V_3 \big| \Big( \tilde c_{-3}-\frac{64}{729 } \tilde
c_1-\frac{5504 }{137781} \tilde c_3-\frac{13568
   }{885735} \tilde c_5
+\cdots \Big)^{(2)} +
   \nonumber \\
\label{conservcc04} &+\big\langle V_3 \big| \Big( \frac{32}{729 }
\tilde c_1-\frac{64}{729 \sqrt{3}} \tilde c_2-\frac{6272 }{19683}
\tilde c_3-\frac{20480 }{19683
   \sqrt{3}} \tilde c_4-\frac{833024 }{885735} \tilde c_5-\frac{3276800 }{1594323 \sqrt{3}}\tilde c_6+\cdots \Big)^{(3)}
   .
\end{align}
Due to the cyclicity of the string vertex, analogous identities
hold by cycling the punctures, (1) $\rightarrow$ (2), (2)
$\rightarrow$ (3), (3) $\rightarrow$ (1). In the next section, we
will give some illustrations of the use of these conservation laws
for the calculation of the three point interaction vertex in open
bosonic string field theory.

\section{Tachyon potentials}
As an application of the conservation laws derived in the previous
section, we will compute here the open string field action
relevant to the tachyon condensation and in order to present not
only an illustration but also an additional information, we
evaluate the action without imposing a gauge choice.

\subsection{Three point correlation function}
Let us begin by introducing some notations. Given three operators
defined in the sliver frame, we write the corresponding states by
$\tilde \phi_1 | 0\rangle$, $\tilde \phi_2 | 0\rangle$ and $\tilde
\phi_3 | 0\rangle$. The three point correlation function involving
these operators will be denoted by
\begin{eqnarray}
\label{three011} \big\langle \tilde \phi_1 , \tilde \phi_2 *
\tilde \phi_3 \big\rangle=\big\langle \tilde \phi_1 , \tilde
\phi_2 , \tilde \phi_3 \big\rangle = \big\langle V_3 \big| \tilde
\phi_1^{(1)} \tilde \phi_2^{(2)} \tilde \phi_3^{(3)} \big| 0 \big
\rangle_{(3)} \otimes \big| 0 \big \rangle_{(2)} \otimes \big| 0
\big \rangle_{(1)} .
\end{eqnarray}
The basic three point correlation function that we will need is
\begin{eqnarray}
\label{three012} \big\langle \tilde c_1 , \tilde c_1 , \tilde c_1
\big\rangle = \big\langle V_3 \big| \tilde c_1^{(1)} \tilde
c_1^{(2)} \tilde c_1^{(3)} \big| 0 \big \rangle_{(3)} \otimes
\big| 0 \big \rangle_{(2)} \otimes \big| 0 \big \rangle_{(1)} .
\end{eqnarray}
To compute this correlator, we use equation (\ref{tr333})
\begin{eqnarray}
\label{fr1} \big\langle \tilde c_1 , \tilde c_1 , \tilde c_1
\big\rangle &=& \big\langle c(\frac{3 \pi}{4}) c(\frac{\pi}{4})
c(-\frac{\pi}{4})\big\rangle_{C_{ \frac{3 \pi}{2}   }} \nonumber
\\
&=& \frac{3^3}{2^3}  \sin^2 \left(\frac{\pi }{3}\right) \sin
\left(\frac{2 \pi }{3}\right) = \frac{3^4 \sqrt{3}}{2^6} \, ,
\end{eqnarray}
where we have used the expression for the correlator $\langle
c(x_1)c(x_2)c(x_3)\rangle_{C_L}$ given in \cite{Okawa:2006vm}
\begin{eqnarray}
\label{ape03}\langle c(x_1)c(x_2)c(x_3)\rangle_{C_L} =
\frac{L^3}{\pi ^3} \sin \left(\frac{\pi
\left(x_1-x_2\right)}{L}\right) \sin
   \left(\frac{\pi  \left(x_1-x_3\right)}{L}\right) \sin \left(\frac{\pi
   \left(x_2-x_3\right)}{L}\right) \, .
\end{eqnarray}
The conservation laws derived in the previous section allow the
computation of all necessary three point functions in terms of the
basic correlator (\ref{fr1}).

Since we will analyze the tachyon potential by performing
computations in the $\mathcal{L}_0$ level truncation, we are going
to define the level of a state as the eigenvalue of the operator
$N=\mathcal{L}_0+1$. This definition is adjusted so that the zero
momentum tachyon defined by the state $\tilde c_1 |0\rangle$ is at
level zero.

Having defined the level number of states contained in the level
expansion of the string field, level of each term in the action is
also defined to be the sum of the levels of the fields involved.
For instance, if states $\tilde \phi_1$, $\tilde \phi_2$, $\tilde
\phi_3$ have level $N_1$, $N_2$, $N_3$ respectively, we assign
level $N_1+N_2+N_3$ to the interaction term
$\langle\tilde\phi_1,\tilde\phi_2,\tilde\phi_3\rangle$. When we
say level $(M,N)$, we mean that the string field includes all
terms with level $\leq M$ while the action includes all terms with
level $\leq N$.

The string field $\Psi$ expanded in the sliver
$\mathcal{L}_0$-basis reads as follows
\begin{eqnarray}
\Psi= \sum_{i=0}^{\infty} x_{i} | \psi^i \rangle \, ,
\end{eqnarray}
where the state $| \psi^i \rangle$ is built by applying the modes
of the $\tilde c(z)$ ghost and the operators $({\cal L}_0+{\cal
L}_0^\dag)^n \equiv \hat{\mathcal{L}}^n$, ${\cal B}_0+{\cal
B}_0^\dag \equiv \hat{\mathcal{B}}$ on the $SL(2,\mathbb{R})$
invariant vacuum $|0\rangle$. The first term in the expansion is
given by the zero-momentum tachyon $| \psi^0 \rangle =\tilde c_1
|0\rangle $. We will restrict our attention to an even-twist and
ghost-number one string field $\Psi$. The tachyon potential we
want to evaluate is defined as
\begin{eqnarray}
\label{norpoten1} V=2 \pi^2 \Big[\frac{1}{2} \langle \Psi,Q_B \Psi
\rangle +\frac{1}{3} \langle \Psi,\Psi*\Psi \rangle  \Big] \, .
\end{eqnarray}

\subsection{The tachyon potential in arbitrary gauge}
In order to illustrate the use of conservation laws, let us
compute in detail the level (2,6) tachyon potential in arbitrary
gauge. Expanding the string field up to level two states, we have
\begin{eqnarray}
\label{psil2} \Psi= x_0 \tilde c_1  |0\rangle + x_1
\hat{\mathcal{L}} \tilde c_1 |0\rangle + x_2 \hat{\mathcal{B}}
\tilde c_0 \tilde c_1 |0\rangle +x_3 \tilde c_{-1}  |0\rangle +
x_4\hat{\mathcal{L}}^2 \tilde c_1 |0\rangle +x_5 \hat{\mathcal{B}}
\hat{\mathcal{L}} \tilde c_0 \tilde c_1 |0\rangle .
\end{eqnarray}

The evaluation of the kinetic term $V_{\text{kin}} = \frac{1}{2}
\langle \Psi, Q_B \Psi \rangle$ requires the action of the BRST
operator and the computation of the BPZ inner product. For the
string field given in equation (\ref{psil2}), to the required
level the computation of the kinetic term is straightforward, we
find
\begin{eqnarray}
\label{kinec0021}V_{\text{kin}} = -\frac{x_0^2}{2}+2 x_1 x_0-2 x_2
x_0+x_3 x_0-2 x_4 x_0+2 x_5 x_0-x_1^2-x_2^2-x_3^2+2 x_1 x_2 .
\end{eqnarray}

The real task is the evaluation of the cubic interaction term
$V_{\text{inter}}=\frac{1}{3} \langle \Psi,\Psi*\Psi \rangle$.
Using the conservation laws derived in the previous section, we
will illustrate the computation of the various cubic couplings.
For instance, plugging (\ref{psil2}) into the definition of the
cubic interaction term, we get a coupling like
\begin{eqnarray}
\label{cou01} \big\langle \tilde c_1 , \tilde c_{-1} , \tilde c_1
\big\rangle = \big\langle V_3 \big| \tilde c_1^{(1)} \tilde
c_{-1}^{(2)} \tilde c_1^{(3)} \big| 0 \big \rangle_{(3)} \otimes
\big| 0 \big \rangle_{(2)} \otimes \big| 0 \big \rangle_{(1)} ,
\end{eqnarray}
where the subscripts denote the labels distinguishing the three
state spaces. We want to relate this term (\ref{cou01}) to the
correlator given in equation (\ref{fr1}). To do that we can simply
use the conservation law for $\tilde c_{-1}$ which is given in
equation (\ref{conservcc02}). Note that the only term that
contributes is described by the replacement $\tilde c^{(2)}_{-1}
\rightarrow -\frac{16}{27} \tilde c^{(2)}_{1}$. Taking into
account these considerations, we obtain
\begin{eqnarray}
\label{cou02} \big\langle \tilde c_1 , \tilde c_{-1} , \tilde c_1
\big\rangle = -\frac{16}{27} \big\langle \tilde c_1 , \tilde c_{1}
, \tilde c_1 \big\rangle = -\frac{3 \sqrt{3}}{4}.
\end{eqnarray}

Following the same procedure, using conservation laws one readily
computes some additional terms
\begin{eqnarray}
\label{cou03} \big\langle \tilde c_{-1} , \tilde c_{-1} , \tilde
c_{1} \big\rangle &=& \frac{64}{243} \big\langle \tilde c_1 ,
\tilde c_{1} , \tilde c_1 \big\rangle = \frac{1}{\sqrt{3}} \, , \\
\label{cou04} \big\langle \tilde c_{-1} , \tilde c_{-1} , \tilde
c_{-1} \big\rangle &=&  0 \, .
\end{eqnarray}

Let us now consider some correlators involving the operators
$\hat{\mathcal{B}}$ and $\hat{\mathcal{L}}$. For example, to
compute the coefficient in front of the $x_0^2 x_5$ interaction we
write
\begin{eqnarray}
\label{cou05} \big\langle \hat{\mathcal{B}} \hat{\mathcal{L}}
\tilde c_0 \tilde c_1 , \tilde c_{1} , \tilde c_{1} \big\rangle
&=& \big\langle V_3 \big| \hat{\mathcal{B}}^{(1)}
\hat{\mathcal{L}}^{(1)} \tilde c_0^{(1)} \tilde c_1^{(1)} \tilde
c_{1}^{(2)} \tilde c_1^{(3)} \big| 0 \big \rangle_{(3)} \otimes
\big| 0 \big \rangle_{(2)} \otimes \big| 0 \big \rangle_{(1)} \, .
\end{eqnarray}
Using the conservation laws (\ref{conservxx1}),
(\ref{conserv00007}) and the following set of commutator and
anticommutator relations
\begin{align}
[\mathcal{B}_0^{(i)},\hat{\mathcal{L}}^{(j)}]&=\delta^{ij}\hat{\mathcal{B}}^{(j)},
\;\;\;\;\;\; [\mathcal{B}_{-1}^{(i)},\hat{\mathcal{L}}^{(j)}]=0, \\
\{\mathcal{B}_0^{(i)},\tilde c_p^{(j)}\}&=\delta^{ij}\delta_{p,0},
\;\;\;\;\;\;
 \{\mathcal{B}_{-1}^{(i)},\tilde
c_p^{(j)}\}=\delta^{ij}\delta_{p,1}, \\
[\mathcal{L}_0^{(i)},\tilde c_p^{(j)}]&=-\delta^{ij} p \, \tilde
c_p^{(j)}, \;\;\; [\mathcal{L}_{-1}^{(i)},\tilde
c_p^{(j)}]=\delta^{ij} (2-p) \, \tilde c_{p-1}^{(j)},
\end{align}
from equation (\ref{cou05}) we obtain
\begin{align}
\label{cou06} \big\langle \hat{\mathcal{B}} \hat{\mathcal{L}}
\tilde c_0 \tilde c_1 , \tilde c_{1} , \tilde c_{1} \big\rangle =
-\frac{8}{9}\big\langle \tilde c_1 , \tilde c_{1} , \tilde c_1
\big\rangle -\frac{2}{9}\pi \big\langle \tilde c_0 \tilde c_1 ,
\tilde c_{1} , \textbf{1} \big\rangle + \frac{1}{18} \pi ^2
\big\langle \tilde c_0 \tilde c_1 , \tilde c_{0} , \textbf{1}
\big\rangle \, ,
\end{align}
where the $\textbf{1}$ appearing in (\ref{cou06}) represents the
$SL(2,\mathbb{R})$ invariant vacuum $|0\rangle$ without any
insertion. In order to write this last equation (\ref{cou06}) in
terms of the basic correlator (\ref{fr1}), we employ the ghost
conservation law (\ref{conservcc01})
\begin{align}
\label{cou07} \big\langle \hat{\mathcal{B}} \hat{\mathcal{L}}
\tilde c_0 \tilde c_1 , \tilde c_{1} , \tilde c_{1} \big\rangle =
\Big( -\frac{8}{9}-\frac{8 \pi }{27 \sqrt{3}}-\frac{8 \pi
^2}{243}\Big) \big\langle \tilde c_1 , \tilde c_{1} , \tilde c_1
\big\rangle = -\frac{9 \sqrt{3}}{8}-\frac{3 \pi }{8}-\frac{\pi
^2}{8 \sqrt{3}} \, .
\end{align}

We see that in this alternative method, the computation of three
point correlation functions avoids the necessity of evaluating any
triple contour integral, the conservation laws allow the
evaluation of all necessary three point functions in terms of the
basic correlator (\ref{fr1}). All the remaining interaction terms,
which come from plugging the string field (\ref{psil2}) into the
definition of the cubic interaction term
$V_{\text{inter}}=\frac{1}{3} \langle \Psi,\Psi*\Psi \rangle$, are
computed similarly. As a result of our computations, we find that
the interacting part of the potential is given by
\begin{align}
\label{cubic0021}V_{\text{inter}} &= \frac{27 \sqrt{3}}{64} x_0^3
-\frac{81 \sqrt{3}}{32} x_0^2 x_1+ \big( \frac{27
\sqrt{3}}{8}+\frac{\sqrt{3} \pi ^2}{16} \big) x_1^2 x_0 + \big(
-\frac{3 \sqrt{3}}{4}-\frac{\pi ^2}{8 \sqrt{3}}+\frac{\pi ^3}{108}
\big)x_1^3 \nonumber \\
&+ \big( \frac{27 \sqrt{3}}{32}+\frac{9 \pi }{16}\big)x_0^2 x_2 +
\big( -\frac{9 \sqrt{3}}{4}-\frac{15 \pi }{8}+\frac{\pi ^2}{8
\sqrt{3}} \big) x_0 x_1 x_2 + \frac{\sqrt{3} \pi ^2}{8}
x_2^3-\frac{3
\sqrt{3}}{4} x_0^2 x_3  \nonumber \\
&+ \big(\frac{9 \pi }{8}+\frac{\sqrt{3} \pi ^2}{16} \big) x_0
x_2^2 + \big(\frac{\pi ^2}{8 \sqrt{3}}-\frac{3 \pi }{4}-\frac{\pi
^3}{36} \big) x_1 x_2^2 + \big( \frac{3 \sqrt{3}}{4}+\frac{3 \pi
}{4}-\frac{\sqrt{3} \pi ^2}{8}+\frac{\pi ^3}{54}\big) x_1^2
x_2 \nonumber \\
&+ \big(\sqrt{3}+\frac{\pi }{6} \big)x_0 x_1 x_3 -\frac{2 \pi
^2}{9 \sqrt{3}} x_1^2 x_3 + \big(-\sqrt{3}-\frac{\pi }{6}\big) x_0
x_2 x_3 -\frac{2 \pi ^2}{9 \sqrt{3}} x_1 x_2 x_3 +
\frac{1}{\sqrt{3}} x_0 x_3^2 \nonumber \\
&+ \big( \frac{2}{3 \sqrt{3}}-\frac{4 \pi }{27} \big) x_1 x_3^2 +
\frac{2}{3 \sqrt{3}} x_2 x_3^2 + \big(\frac{27
\sqrt{3}}{8}-\frac{\sqrt{3} \pi ^2}{8} \big) x_4 x_0^2 +
\big(-\frac{9 \sqrt{3}}{2}-\frac{\pi ^3}{36} \big) x_0 x_1 x_4 \nonumber \\
&+ \big( \frac{3 \sqrt{3}}{2}+\frac{3 \pi }{2}-\frac{5 \pi
^3}{108} \big) x_0 x_2 x_4 + \frac{\pi ^4}{81 \sqrt{3}} x_1 x_2
x_4 -\frac{\pi ^4}{27 \sqrt{3}} x_2^2 x_4 + \frac{5 \pi ^2}{9
\sqrt{3}} x_0 x_3 x_4 \nonumber \\
&+ \big(-\frac{2 \pi ^2}{9 \sqrt{3}}+\frac{4 \pi ^3}{243} \big)
x_1 x_3 x_4 + \big( \frac{8 \pi ^3}{243}-\frac{2 \pi ^2}{9
\sqrt{3}}\big) x_2 x_3 x_4 + \big( \frac{8}{9 \sqrt{3}}-\frac{32
\pi }{81}-\frac{16 \pi ^2}{81 \sqrt{3}} \big) x_3^2 x_4 \nonumber \\
&+ \frac{\pi ^4}{36 \sqrt{3}} x_4^2 x_0  + \big(\frac{2 \pi ^4}{81
\sqrt{3}}+\frac{8 \pi^5}{2187}-\frac{4 \pi ^6}{6561 \sqrt{3}}\big) x_4^3+
\big(\frac{32 \pi ^3}{729}-\frac{8 \pi ^2}{27 \sqrt{3}}-\frac{16 \pi ^4}{729 \sqrt{3}} \big) x_4^2 x_3 \nonumber \\
&+\big( -\frac{23 \pi ^4}{486 \sqrt{3}}+\frac{4 \pi ^5}{2187}\big)
x_4^2 x_2 + \big( \frac{\pi ^4}{54 \sqrt{3}}+\frac{\pi
^5}{729}\big) x_4^2 x_1 + \big( -\frac{9 \sqrt{3}}{8}-\frac{3 \pi
}{8}-\frac{\pi ^2}{8 \sqrt{3}}\big) x_0^2 x_5 \nonumber \\
&+ \big( \frac{3 \sqrt{3}}{2}+\frac{3 \pi }{4}+\frac{\pi ^2}{2
\sqrt{3}}+\frac{\pi ^3}{108}\big)x_0 x_1 x_5 -\frac{\pi ^4}{81
\sqrt{3}} x_1^2 x_5 + \big(\frac{\pi ^3}{36} -\frac{3
\pi}{2}-\frac{\pi ^2}{2
\sqrt{3}} \big)x_0 x_2 x_5 \nonumber \\
&+\frac{\pi ^4}{27 \sqrt{3}} x_1 x_2 x_5 + \frac{\pi ^2}{9
\sqrt{3}} x_0 x_3 x_5 + \big( -\frac{4 \pi ^2}{9 \sqrt{3}}-\frac{4
\pi ^3}{243} \big) x_1 x_3 x_5 + \big( \frac{8}{9
\sqrt{3}}-\frac{8 \pi }{81} \big) x_3^2 x_5 \nonumber \\
&+ \frac{5 \pi ^4}{162 \sqrt{3}} x_0 x_4 x_5 +  \big( -\frac{2 \pi
^4}{243 \sqrt{3}}+\frac{2 \pi ^5}{2187} \big) x_1 x_4 x_5 + \big(
-\frac{4 \pi ^4}{81 \sqrt{3}}-\frac{4 \pi ^5}{729} \big) x_2 x_4
x_5 \nonumber \\
&+ \big( -\frac{32 \pi ^2}{27 \sqrt{3}}+\frac{8 \pi
^3}{729}-\frac{16 \pi ^4}{729 \sqrt{3}} \big) x_3 x_4 x_5 +  \big(
-\frac{2 \pi ^4}{243 \sqrt{3}}-\frac{16 \pi ^5}{2187}+\frac{4 \pi
^6}{2187 \sqrt{3}}
\big) x_4 x_5^2 \nonumber \\
&-\frac{\pi ^4}{108 \sqrt{3}} x_0 x_5^2 + \big( \frac{5 \pi
^4}{162 \sqrt{3}}+\frac{\pi ^5}{729} \big) x_1 x_5^2 + \big(
-\frac{22 \pi ^4}{243 \sqrt{3}}+\frac{2 \pi ^5}{2187}-\frac{8 \pi
^6}{6561 \sqrt{3}}
\big) x_4^2 x_5 \nonumber \\
&+ \frac{\pi ^4}{18 \sqrt{3}} x_2 x_5^2 + \big( \frac{2 \pi ^4}{27
\sqrt{3}}+\frac{2 \pi ^5}{729} \big) x_5^3 \, .
\end{align}

Once we have obtained the kinetic (\ref{kinec0021}) and the cubic
interaction term (\ref{cubic0021}), we can try to find a critical
point of the level (2,6) tachyon potential $V= 2 \pi^2
(V_{\text{kin}}+V_{\text{inter}})$ without using any gauge. Since
in the complete theory there is an infinite-dimensional gauge
orbit of equivalent locally stable vacua, we would expect
undetermined parameters at the critical value of the potential.
Nevertheless the breakdown of gauge invariance caused by level
truncation, tell us that the critical points of the
level-truncated tachyon potential have only a discrete set of
solutions. At the level (2,6), for example, we find a definite
critical point. The critical point is found at $x_0 = 0.218563$,
$x_1 = -0.009950$, $x_2 = 0.162003$, $x_3 = 0.053121$, $x_4 =
0.027101$, $x_5 = -0.063514$, and gives the value of $V=-1.049082$
which is $4.90 \%$ greater than the required vacuum energy. In
contrast, in the Schnabl gauge \cite{AldoArroyo:2009hf}, we have
obtained $V_{\text{Sch}}=-1.046622$ which is $4.66 \%$ greater
than the required vacuum energy (with $x_0 = 0.702361$, $x_1 =
0.165917$, $x_2 = 0.165917$, $x_3 = 0.036787$, $x_4 = 0.044922$,
$x_5 =0.089844$).

We could continue performing higher level computations, since
these computations follow the same procedures previously shown, at
this stage we only want to present our results. At the next level,
namely expanding the string field up to level three states
\begin{align}
\label{psil33} \Psi &= x_0 \tilde c_1  |0\rangle + x_1
\hat{\mathcal{L}} \tilde c_1 |0\rangle + x_2 \hat{\mathcal{B}}
\tilde c_0 \tilde c_1 |0\rangle +x_3 \tilde c_{-1}  |0\rangle +
x_4\hat{\mathcal{L}}^2 \tilde c_1 |0\rangle +x_5 \hat{\mathcal{B}}
\hat{\mathcal{L}} \tilde c_0 \tilde c_1 |0\rangle \nonumber \\
&+x_6 \hat{\mathcal{L}} \tilde c_{-1} |0\rangle + x_7
\hat{\mathcal{L}}^3 \tilde c_{1} |0\rangle + x_8 \hat{\mathcal{B}}
\tilde c_{-2} \tilde c_1 |0\rangle + x_9 \hat{\mathcal{B}} \tilde
c_{-1} \tilde c_0 |0\rangle + x_{10} \hat{\mathcal{B}}
\hat{\mathcal{L}}^2 \tilde c_{0} \tilde c_1 |0\rangle \, ,
\end{align}
we have found that there are five critical points for the level
(3,9) tachyon potential. The first of these critical points is
located at $x_0 = 0.320696$, $x_1 =-0.011758$, $x_2 = 0.142364$,
$x_3 = 0.095307$, $x_4 = -0.033742$, $x_5 = -0.032490$,
$x_6=-0.048882$, $x_7=0.014538$, $x_8=-0.095482$, $x_9=-0.016128$,
$x_{10}=0.008261$ and gives the value of $V=-0.999705$ which is
$99.97 \%$ of the required vacuum energy. The second point is at
$x_0 = 0.375516$, $x_1 =0.016889$, $x_2 = 0.145665$, $x_3
=0.089377$, $x_4 =-0.024954$, $x_5 =-0.013322$, $x_6=-0.036865$,
$x_7=0.010329$, $x_8=-0.072036$, $x_9=-0.009889$,
$x_{10}=0.003576$ and gives the value of $V=-0.999533$ which is
$99.95 \%$ of the required vacuum energy. The third point is at
$x_0 = 0.465295$, $x_1 =0.038118$, $x_2 = 0.124101$, $x_3
=0.038559$, $x_4 =-0.019113$, $x_5 =0.015616$, $x_6=-0.026136$,
$x_7=-0.002337$, $x_8=-0.066607$, $x_9=-0.013154$,
$x_{10}=-0.008964$ and gives the value of $V=-1.001516$ which is
$100.15 \%$ of the required vacuum energy. The fourth point is at
$x_0 = 0.670369$, $x_1 =0.189001$, $x_2 = 0.171491$, $x_3
=0.103166$, $x_4 =0.023686$, $x_5 =0.029369$, $x_6=-0.000273$,
$x_7=0.001751$, $x_8=0.020806$, $x_9=0.022997$, $x_{10}=0.001284$
and gives the value of $V=-0.999674$ which is $99.96 \%$ of the
required vacuum energy. The last point is at $x_0 = 0.673922$,
$x_1 =-0.072591$, $x_2 = -0.095384$, $x_3 =-0.246487$, $x_4
=-0.041749$, $x_5 =0.160009$, $x_6=-0.027547$, $x_7=-0.006767$,
$x_8=-0.300267$, $x_9=-0.062356$, $x_{10}=-0.044929$ and gives the
value of $V=-1.009647$ which is $100.96 \%$ of the required vacuum
energy.

Up to the level that we have explored with our computations, we
have noticed that as the level of truncation is increased, the
multiplicity of the candidate solutions continues to grow. While
some solutions approach the correct value, others do not, so
without some further criterion for selecting solutions, it does
not seem possible to isolate a good candidate for the vacuum in
the level-truncated, non-gauge-fixed theory. This difficulty in
solving the theory without gauge fixing clearly arises from the
presence of a continuous family of gauge equivalent vacua in the
full theory. A unique solution of the tachyon potential which
represents the stable vacuum at each level has the property that
it can be determined from the branch of the gauge fixed effective
potential connecting the perturbative and nonperturbative vacua,
as it was done in \cite{AldoArroyo:2009hf} for the tachyon
potential in the Schnabl gauge.

\section{Summary and discussion}
For a string field expanded in the sliver $\mathcal{L}_0$-basis,
we have provided an efficient and alternative method, based on
conservation laws, to compute the three point interaction vertex
of the open bosonic string field theory action. As we have seen,
by some explicit computations, these conservation laws are easy to
apply for low level by hand calculations and they can be
implemented for high level computer calculations.

As an application of the conservation laws found in this paper,
using the $\mathcal{L}_0$ level truncation scheme, we have
computed the open bosonic string field action relevant to the
tachyon condensation. We have explored the possibility of
determining the locally stable vacuum without gauge fixing. At
higher levels we have found that there are many candidates for the
stable vacuum, the multiplicity of these solutions is due to the
presence of a continuous family of gauge equivalent vacua in the
full theory.

A problem that should be analyzed employing the methods developed
in this work would be the evaluation of the tachyon potential in
the sliver frame for the case of the modified cubic superstring
field theory. The shape of the effective potential in this theory
was already conjectured by Erler \cite{Erler:2007xt}. This issue
is very puzzling since for the case of the modified cubic
superstring field theory, the tachyon has vanishing expectation
value at the local minimum of the effective potential, so the
tachyon vacuum sits directly below the perturbative vacuum.

The procedures developed in our work could be applicable to
computations in Berkovits superstring field theory as well. The
relevant string field theory is non-polynomial
\cite{Berkovits:1995ab}, but since the theory is based on Witten's
associative star product, the methods discussed in this paper
would apply with minor modifications. For instance, we should
extend the conservation laws for the case of the $n$-point
interaction vertex which involves the gluing of $n$-strings with
$n \geq 2$.

\section*{Acknowledgements}
I would like to thank Ted Erler, Michael Kiermaier, Michael
Kroyter and Barton Zwiebach for useful discussions. This work is
supported by FAPESP grant 2010/05990-0.






\end{document}